\documentclass{pinchcr}   
                                             
\usepackage{color}
\usepackage[english]{babel}
\usepackage{comment}
\usepackage{setspace}
\usepackage{epsfig,psfrag}
\usepackage{amssymb,amsmath}
\usepackage{nicefrac}
\usepackage{bm}

\renewcommand{\i}{\mathrm{i}}

\renewcommand{\vector}[1]{\mathbf{#1}}

\newcommand{\weg}[1]{}

\newcommand{\del}{\partial}

\newcommand{\eq}[1]{(\ref{eq:#1})}
\newcommand{\Eq}[1]{Eq.~(\ref{eq:#1})}

\newcommand{\Fig}[1]{Fig.~\ref{fig:#1}}

\newcommand{\Sect}[1]{Sect.~\ref{sec:#1}}

\renewcommand{\vec}{\mathbf}
\newcommand{\ddf}[1]{\mathrm{d}^{d}\! #1 \,}
\renewcommand{\v}{\mathrm{V}}
\renewcommand{\a}{\mathrm{A}}

\renewcommand*\thesection{\arabic{section}}
\renewcommand*\thesubsection{\thesection.\arabic{subsection}}
\renewcommand*\thesubsubsection{\thesubsection.\arabic{subsubsection}}
\renewcommand{\thetable}{\arabic{table}}  
\renewcommand{\thefigure}{\arabic{figure}}

\begin{document}

\maintext

\chapter*{Non-thermal fixed points: 
universality, topology, \& turbulence
 in Bose gases}

{\sc B. Nowak, S. Erne, M. Karl, J. Schole, D. Sexty,  
and T. Gasenzer\footnote{Prepared for the proceedings of the Summer school: Strongly interacting quantum systems out of equilibrium, held
30 July--24 August 2012 at Ecole de Physique des Houches, Les Houches, France, to which T.G. contributed a seminar talk.}}\\

Institut f\"ur Theoretische Physik der Ruprecht-Karls-Universit\"at Heidelberg,
 
Philosophenweg 16, 69120 Heidelberg, Germany, and

ExtreMe Matter Institute EMMI, GSI Helmholtzzentrum f\"ur Schwerionen- 

forschung GmbH, Planckstra\ss e~1, 64291~Darmstadt, Germany

\section{Introduction}

At a second-order phase transition different types of order of a physical system meet one another, giving rise to universal critical properties which are independent of the microscopic details of the system.
This kind of universality is an extremely successful concept in characterising equilibrium states of matter and classifying different phenomena in terms of just a few classes governed by the same critical properties. 
The appearance of an ordered state which does not possess the full symmetry deriving from the conservation laws obeyed by the system's dynamics is called spontaneous symmetry breaking.
The order induced by this symmetry breaking has long been known to allow for the appearance of (quasi-)topological defects such as solitons or vortices, objects we are particularly interested in here.

In these notes we leave equilibrium systems aside and aim at sketching a picture of related concepts far from thermal equilibrium.
We consider the example of nearly coherent Bose gases brought far out of equilibrium and discuss their behaviour in view of connections between universal properties, (quasi-)topological field configurations and turbulent dynamics.
We demonstrate that the isolated Bose gas, on its way back to thermal equilibrium, can approach metastable non-equilibrium configurations and spend a long time in their vicinity.
In such configurations, which have been termed \textit{non-thermal fixed points}, the system shows universal long-range properties manifest through scaling, i.e., self-similar correlations.
The time evolution near such fixed points is demonstrated to undergo critical slowing down.
The spatial field pattern, at the same time, is characterized by the appearance of defects and domain formation whose geometry gives rise to the particular scaling laws seen in the correlation functions.
We obtain an overall picture which connects well-known concepts for describing universal dynamics such as wave-turbulence, superfluid turbulence, and (quasi-)topological excitations.
This allows to bring together an excitingly wide range of concepts and methods with an excitingly wide spectrum of applications.
Beyond the immediate implications for simple low-energy degenerate quantum gases, phenomena such as topological configurations in solids, in soft matter, the dynamics of the quark-gluon plasma created in heavy-ion collisions, or the reheating of the post-inflationary universe come in sight.
Vice versa, an ultracold quantum gas offers itself as a kind of the often discussed `quantum simulator' for universal dynamics of systems which are more difficult to access experimentally, like some of the above.
Not the least, the extension to fermionic and gauge fields should bear many new interesting aspects.

Time-evolution far from equilibrium of a system with many degrees of freedom is often characterized by the appearance of widely different scales.
Usually, a period of fast motion is followed by a much longer period of much slower motion. 
In the same way a separation of spatial scales allows descriptions in terms of statistical concepts such as hydrodynamics or renormalization-group theory.
Taking into account the well-developed concepts and concrete results for equilibrium and near-equilibrium systems (cf., e.g., \shortciteANP{Hohenberg1977a}, \citeyearNP{Hohenberg1977a}), it has to be expected that far-from-equilibrium time evolution will show universal behaviour, with fixed points or partial fixed points being just a small subclass of the possible phenomena.
Recently, the discussion of the possible things waiting to be discovered has increased in intensity, in particular in the context of possible `prethermalisation' phenomena \shortcite{Berges:2004ce,Bonini1999a,Aarts2000a,Gasenzer:2005ze,Berges:2007ym,Barnett2011a,Kitagawa2011a,Gring2011a,Kollar2011a}. 
Such dynamics could be considered to hint to a wider class of universal phenomena which will allow to classify and more deeply understand non-equilibrium physics.

\section{Strong wave turbulence and vortical flow}

In analogy to equilibrium phase transitions and criticality in driven systems it has been proposed that transient stationarity arising in the time evolution of an initially strongly perturbed quantum many-body system can reflect the existence of \textit{non-thermal fixed points} \shortcite{Berges:2008wm}. 
At the same time, these fixed points were numerically demonstrated to occur in the field evolution imposed by an $O(N)$-symmetric, relativistic, non-linear scalar model:
By means of parametrically resonant oscillations of the field expectation value, and non-linear amplification, a broad range of modes can be excited initially.
While falling back to equilibrium their occupation number spectra shows scaling behaviour.
Remarkably, the respective power-law exponents confirmed analytical predictions in the infrared domain of long-wavelength excitations where standard descriptions in terms of Boltzmann-type kinetic equations break down.
Found when searching for stationary scaling solutions of non-perturbative dynamic Dyson equations, with the help of Zakharov integral transformations known in wave-turbulence theory \shortcite{Zakharov1992a,Nazarenko2011a}, these infrared power laws are interpreted to constitute the previously mysterious strong wave turbulence which had been considered out of reach of kinetic theory \shortcite{Scheppach:2009wu}.

In the following we identify this  strong wave turbulence, considering the example of a non-relativistic, superfluid Bose gas in two spatial dimensions, with the appearance of \textit{quantised vortices}.
On scales considerably smaller than the mean distance between vortices, the velocity field $v(r)$ associated with the rotational flow decays as $1/r$ with growing distance $r$ from the nearest vortex core. 
As a result, the angle-averaged kinetic energy distribution $\sim m v^{2}/2$ of Bosons gives rise to the single-particle momentum spectrum $n(k)\sim k^{-4}$ identical to that predicted for strong wave turbulence by  \shortciteN{Scheppach:2009wu}.
Hence, the non-thermal fixed point appears to correspond to a configuration bearing a dilute ensemble of vortices. 
In this section we give evidence for this interpretation which allows to conjecture a deep link between the extended kinetic-theory picture of wave turbulence on the one side and the theory of non-linear (quasi-)topological field configurations and superfluid turbulence on the other.

\subsection{Non-thermal fixed points and weak wave turbulence}

Most generally, a non-thermal fixed point can be defined as a metastable state of a many-body system~\shortcite{Bonini1999a,Aarts2000a,Berges:2008wm}.
While one may include open systems into the discussion, e.g. driven ones and those experiencing dissipation \shortcite{Diehl2008b}, we will, in the following, restrict ourselves to closed systems. 
In analogy to fixed points in scaling flows, one furthermore considers in particular solutions with power-law behaviour of correlation functions~\shortcite{Berges:2008wm,Berges:2008sr,Scheppach:2009wu}. 
While, precisely at the fixed point, the system is metastable and characterised by scaling correlation functions in the infrared limit of infinitesimally slow modes, it will vary in time away from the fixed point and show power-law correlations within a finite scaling regime.
From the physics of turbulence, it is well known that such states occur as a consequence of local conservation laws in momentum space.
Such unintuitive locally conserved currents in momentum space instead of position space imply the possibility of gain- and loss-less transport processes between different scales, giving rise to so-called cascades. 
The most prominent example of this kind is fully developed classical fluid turbulence. 
It comprises a quasi-stationary flow of kinetic energy from large to small spatial scales, i.e., from low to high momenta \shortcite{Richardson1920a}. 
The energy is fed in, e.g., by a stirrer, at a large scale and finally dissipated into heat at the microscopic scale defined by the fluid's viscosity. 
The corresponding energy spectrum exhibits the famous Kolmogorov-Obukhov five-third scaling of the radial energy distribution $E(k)\sim k^{-5/3}$~\shortcite{Kolmogorov1941a,Obukhov1941a}.

Considering a dilute Bose gas one has to take into account its compressibility, allowing for collective sound excitations of the particles.
In addition to the density of an incompressible fluid the system is thus characterised by the dispersion relation between momentum and energy of its excitations.
This allows for the techniques of wave turbulence to be invoked when looking into turbulence phenomena of a Bose gas \shortcite{Zakharov1992a,Nazarenko2011a}.
The mathematically best-controlled case is that of weak wave turbulence which rests on the analysis of stationary solutions of Boltzmann-type kinetic equations.
Within a certain range of momenta $k$ and times $t$, for not too strongly excited systems, the quantum Boltzmann equation (QBE)
\begin{align}
  \partial_{t}n(\vec k,t)
  &= I(\vec k,t),
  \label{eq:QKinEq}
  \\
   I(\vec k,t)
  &= \int\ddf{p}\ddf{q}\ddf{r}|T_{\vec k\vec p\vec q\vec r}|^{2}\delta(\vec k+\vec p -\vec q - \vec r)\,
  \delta(\omega_{\vec k}+\omega_{\vec p}-\omega_{\vec q}-\omega_{\vec r})
  \nonumber\\
  &\qquad\quad\times\
  [(n_{\vec k}+1)(n_{\vec p}+1)n_{\vec q}n_{\vec r}-
  n_{\vec k}n_{\vec p}(n_{\vec q}+1)(n_{\vec r}+1)],
  \label{eq:KinScattInt}
\end{align}
well describes the time-evolution of the momentum-mode occupation numbers $n_{\vec k}\equiv n(\mathbf{k},t) = \langle\Phi^{\dagger}(\mathbf{k},t) \Phi(\mathbf{k},t) \rangle$ of an interacting degenerate Bose gas.
Here, $\Phi$ denotes the quantum field operator describing the Bose system, $[\Phi(\vec x,t),\Phi^{\dagger}(\vec y,t)]=\delta(\vec x-\vec y)$, while all other equal-time commutators vanish. 
We consider only two-to-two elastic collisions quantified by the $T$-matrix elements which for dilute, weakly interacting atomic gases reduce to a single quantity, the $s$-wave scattering length $a$, i.e., $|T_{\vec k\vec p\vec q\vec r}|\equiv g= \mathit{const.} \times a$. 

Zeroes of the scattering integral $I(\mathbf{k})$ correspond to fixed points of the time evolution within the regime of applicability of the QBE \shortcite{Zakharov1992a}. 
Most prominent amongst these are the thermal fixed point corresponding to the system in thermal equilibrium, $n_{\vec k}=\{\exp[\beta\omega(\vec k)]-1\}^{-1}$, and the trivial fixed point $n_{\mathbf{k}}=\mathit{const.}$ 
At both fixed points the scattering integral vanishes and $n(\mathbf{k},t)$ becomes independent of $t$.
Note that both, the trivial and the Bose-Einstein distribution (in the Rayleigh-Jeans regime), taking $\omega(\mathbf{k})\sim k^{2}$, show a power-law behaviour of the form $n_{\vec k}\sim k^{-\zeta}$ with $\zeta=0$ and $\zeta=2$, respectively.

The theory of weak wave turbulence \shortcite{Zakharov1992a} allows to analytically derive further, \emph{non-thermal} fixed points at which the occupation number $n_{\mathbf{k}}$ obeys a scaling law of the form $n_{\vec k}\sim k^{-\zeta}$ with, in general, $\zeta\not=2$. 
As in classical turbulence one expects that universal scaling appears within a certain regime of momenta, the inertial range. 
According to this picture, outside the scaling regime excitation quanta enter the system from an external or internal source and/or leave it into a sink, whereas there are no sources and sinks within the inertial interval where the quanta are transported from momentum shell to momentum shell. 
This process is described by a continuity equation in momentum space, with a momentum-independent, radially oriented current vector.\footnote{Note that justification of this assumption, i.e., locality of the transport, needs to be checked for each particular wave-turbulent solution, cf., e.g.,~\shortciteN{Zakharov1992a}.}  
A central aspect of weak-wave-turbulence theory is that the quantum Boltzmann equation can be cast into different such equations~\shortcite{Zakharov1992a}, for the radial densities of particles, $N(k)=(2k)^{d-1}\pi n(k)$, and energy, $E(k)=(2k)^{d-1}\pi \varepsilon(k)$, $\varepsilon(k)=\omega(k)n(k)$,
\begin{align}
  \partial_{t}N(k,t) 
  &= -\partial_{k}Q(k),
  \label{eq:BalEqQ}\\
  \partial_{t}E(k,t) 
  &= -\partial_{k}P(k).
  \label{eq:BalEqP}
  \end{align}
Taking either the radial particle current $Q(k)=(2k)^{d-1}\pi Q_{k}(k)$ or the energy current $P(k)=(2k)^{d-1}\pi P_{k}(k)$ to be independent of $k$, one derives different scaling exponents. 
The resulting exponents\footnote{The superscript UV (ultraviolet) in \Eq{kappaUV} refers to the regime of large momenta where the description in terms of a kinetic equation is expected to be accurate.} are
\begin{align}
   \zeta^\mathrm{UV}_{Q}= d-2/3 ,\quad
   \zeta^\mathrm{UV}_{P}= d.
   \label{eq:kappaUV}
\end{align}
These exponents can be obtained by simple power counting:
Combining Eqs.~\eq{QKinEq} and  \eq{BalEqQ} gives the radial relation $\partial_{k}Q(k)\sim k^{d-1}I(k)$ which implies that stationarity requires $k^{d}I(k)$ to become $k$-independent, i.e., scale as $k^{0}$.
Counting all powers of $k$ in $I(k)$, \Eq{KinScattInt}, in the wave-kinetic regime where the terms of third order in the occupation numbers dominate the scattering integral, this requires  $n(k)\sim k^{-d+2/3}$.
Analogously one infers the exponent $\zeta^\mathrm{UV}_{P}$ from the balance equation \eq{BalEqP} for the energy density $\varepsilon(k)\sim k^{2}n(k)$.
Despite this simple procedure, the existence of the respective scaling solutions has to and can be derived rigorously from the quantum Boltzmann equation by means of Zakharov conformal integral transforms~\shortcite{Zakharov1992a}. 

As we will illustrate for our case in \Sect{Fluxes}, the energy flux generically constitutes a direct cascade to larger $k$, whereas the particle flux corresponds to an inverse cascade.
While the character of the fluxes is entirely determined by the properties of the system, it turns out that there must be at least two sinks, where the particles and the energy can flow to~\shortcite{Zakharov1992a}. 
Let us assume that an external source introduces particles with energy $\omega_{s}$ at some scale $k_{s}$ and that energy and particles leave at scales $k=0$ and $k=K$.
Calling $\Gamma_{s}$, $\Gamma_{0}$, and $\Gamma_{K}$ the injection/ejection rates of particle number at the respective scales, number and energy conservation imply $\Gamma_{s}=\Gamma_{0}+\Gamma_{K}$ and $\omega_{s}\Gamma_{s}=\omega_{0}\Gamma_{0}+\omega_{K}\Gamma_{K}$, respectively.
Inverting these conditions,
\begin{align}
\Gamma_{0}
=\frac{\omega_{K}-\omega_{s}}{\omega_{K}-\omega_{0}}\Gamma_{s},
\qquad
\Gamma_{K}
=\frac{\omega_{s}-\omega_{0}}{\omega_{K}-\omega_{0}}\Gamma_{s},
\end{align}
shows that for $\omega_{0}\ll\omega_{s}\ll\omega_{K}$ the particles are ejected at $\omega_{0}$ whereas energy is dissipated at $\omega_{K}$ \shortcite{Gurarie1995a}.

\subsection{Infrared scaling as strong wave turbulence}
\label{sec:IRScalingAsSWT}

\begin{figure}[t]
\begin{center} 
\includegraphics[width=0.9 \textwidth]{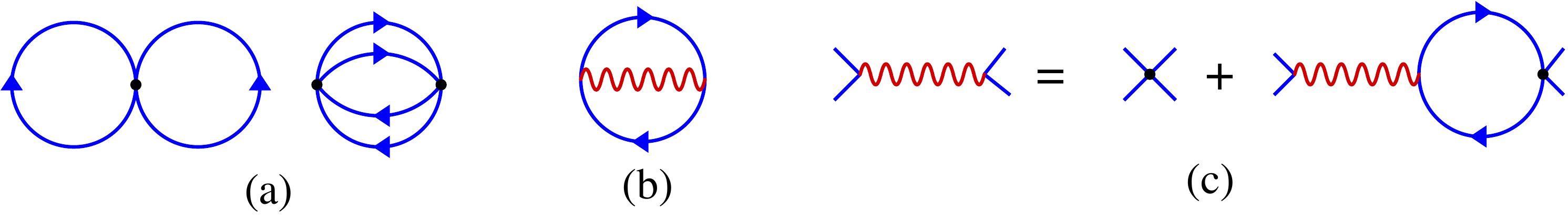}
\caption{2PI diagrams of the loop expansion of $\Gamma_2[G]$.
(a) The two lowest-order diagrams of the loop expansion which lead to the quantum Boltzmann equation. 
Black dots represent the bare vertex $\sim g\delta(x-y)$, solid lines the full propagator $G(x,y)$.
(b) Diagram representing the resummation approximation which, in the IR, replaces the diagrams in (a) and gives rise to the scaling of the $T$-matrix in the IR regime.
(c) The wiggly line is the two-point resummed vertex function which is represented as a sum of bubble-chain diagrams.
\label{fig:2PI}}
\end{center}
\end{figure}
%
Given a positive scaling exponent $\zeta$ momentum occupation numbers $n(k)\sim k^{-\zeta}$ grow large in the IR regime of small $k$.  
Keeping the coupling $g$ fixed, the QBE fails for $n_{\vec k}\gtrsim g^{-1}$ where perturbative contributions to the scattering integral $I(k)$ of order higher than $g^{2}$ are no longer negligible. 
To find scaling solutions in the IR, an approach beyond this perturbative approximation is required\footnote{
For more details we recommend to consult 
\protect\shortciteN{Scheppach:2009wu} for details of the procedure summarised in the following.
}.
 This is available through quantum-field dynamic equations derived from the two-particle irreducible (2PI) effective action or $\Phi$-functional \shortcite{Luttinger1960a,Baym1962a,Cornwall1974a} which can be expanded in terms of 2PI closed loop diagrams the lowest-order ones are sketched in \Fig{2PI}a.
 The solid (blue) lines denote the time-ordered Green's function $G(x,y)=\langle\mathcal{T}\Phi^{\dagger}(x)\Phi(y)\rangle_c$ which, in turn, is a solution of the real-time Dyson equation. 
The Dyson equation is derived from the action by use of Hamilton's variational principle.
It contains a time-evolution equation \eq{QKinEq} for the momentum-mode occupation numbers $n_{\vec k}$. 
As before one considers zeros of the resulting scattering integral which can be expressed, within the 2PI approach, in terms of the self-consistently determined Green's function $G$, connected by the bare scattering vertices of the theory.
The scattering integral \eq{KinScattInt} of the QBE \eq{QKinEq} is recovered within the expansion of the action up to the two-loop diagrams in \Fig{2PI}a.

To describe the IR kinetics one needs to go beyond this approximation.
Resumming an infinite set of loop diagrams contributing to the 2PI effective action \shortcite{Berges:2001fi,Aarts:2002dj}, also, e.g., \shortcite{Berges:2004yj,Gasenzer:2005ze,Gasenzer2009a}
leads to a non-perturbative, effectively renormalised coupling $g_\mathrm{eff}(k)$ in the dynamic equations \shortcite{Berges:2008wm,Berges:2008sr,Scheppach:2009wu}, see \Fig{2PI}b and c. 
In particular, this coupling becomes suppressed in the IR to below its bare value $g$ which, in effect, leads to an even steeper rise of the particle spectrum $n_{\vec k}$ \shortcite{Berges:2008wm}.
The IR scaling exponents for the radial particle and energy flows of a Bose gas in $d$ dimensions which constitute the resulting \emph{strong wave turbulence} were derived by \shortciteN{Scheppach:2009wu} to be
\begin{align}
   \zeta^\mathrm{IR}_{Q}= d+2 ,\quad
\zeta^\mathrm{IR}_{P}= d+2+z \,,
  \label{eq:kappaIR}
\end{align}
where $z$ is the dynamical scaling exponent accounting for the scaling of the dispersion $\omega(sk)=s^{z}\omega(k)$.
From the point of view of its scaling, the $T$-matrix in the scattering integral of the QBE can be replaced by an 
effective many-body $T$-matrix, $T^{\mathrm{eff}}_{\vec k\vec p\vec q\vec r}\equiv T^{\mathrm{eff}}_{\vec k+\vec p,\vec q+\vec r}$.
This effective $T$-matrix is more complex than but scales like
\begin{align}
|T^{\mathrm{eff}}_{\mathbf{k}}|\equiv|T^{\mathrm{eff}}_{\mathbf{k},\mathbf{k}}|\sim |gCk^{z-2}/[1+C'gk^{d-2}n_{\mathbf{k}}]| \,,
  \label{eq:lambdaeff}
\end{align}
$k=|\mathbf{k}|$, where $C'$ is some constant which fine-tunes the position of the transition from UV to IR scaling. 
The second term in the denominator can be related to the validity criterion of the kinetic equation \eq{QKinEq} (cf., e.g., \shortciteANP{Svistunov1991a}, \citeyearNP{Svistunov1991a}) in $d$ dimensions,
\begin{align}
g\int_0^k \mathrm{d}^dk' \, n(\vector{k}') \ll \frac{k^2}{2m}\, .
  \label{eq:kriterion}
\end{align}
For a scaling distribution $n(k)\sim k^{-\zeta}$ this translates into $2\Omega_d mgk^{d-2}n(k)\ll 1$, with $\Omega_d$ the surface of a unit sphere in $d$ dimensions.
For small $n_{\mathbf{k}}$ and $z=2$ one recovers the weak-wave-turbulence case discussed in the previous section.
For large $n_{\mathbf{k}}$, the second term in the denominator dominates which
implies a power-law behaviour $|T^{\mathrm{eff}}_{\mathbf{k}}|^{2}\sim k^{2(\zeta-d+z)}$ and, as a consequence, the modified scaling \eq{kappaIR} of $n_{\mathbf{k}}$ in the infrared regime of small wave numbers.
Moreover, the coupling becomes universal in the sense that it is now independent of $g$ which is cancelled out by the leading denominator term in \eq{lambdaeff}. 

\begin{figure}[t!]
\centering
   \includegraphics[width=0.6\textwidth]{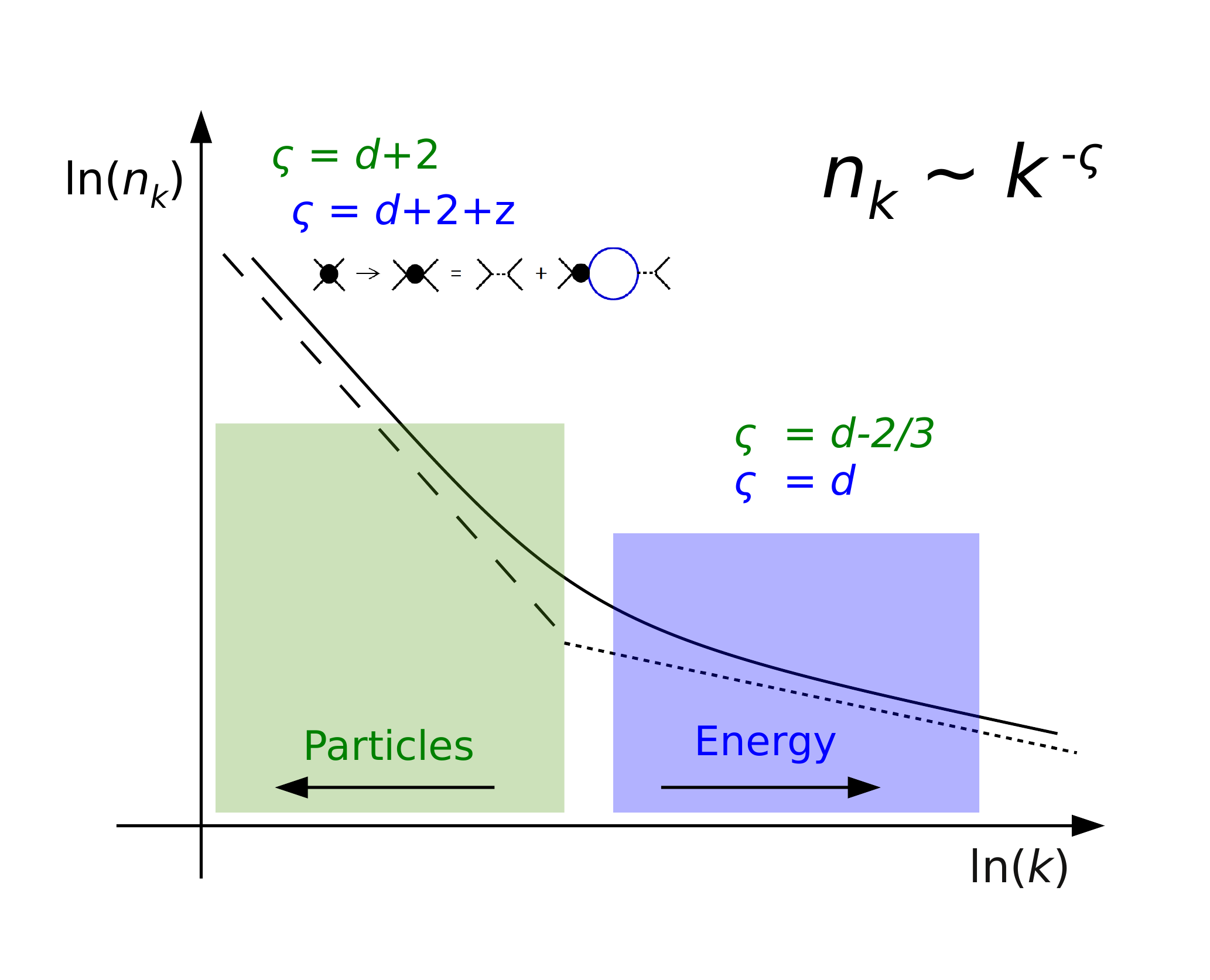}
    \caption{Sketch of the single-particle mode occupation number $n(k)$ of a Bose gas in $d$ dimensions, as a function of the radial momentum $k$ of the system at  the non-thermal fixed point, as predicted by {\protect\shortciteN{Scheppach:2009wu}}. 
    At large momenta, weak wave turbulence scaling is recovered, with the Kolmogorov-Zakharov spectra $\zeta=d-2/3$ (particle flux) and $\zeta=d$ (energy flux). 
    At low momenta, new scaling solutions are predicted, with $\zeta=d+2$ (particle flux) and $\zeta=d+2+z$ (energy flux), termed strong turbulence. 
    Colours indicate the expected character of an IR particle flux and a UV energy flux. }
    \label{fig:NTFP}
    \end{figure}

In \Fig{NTFP}, we summarise the non-thermal fixed point scaling predicted with the loop-resummed 2PI effective action, for a dilute Bose gas in $d$ dimensions.
%
Direct confirmation of the scaling  \Eq{kappaIR} by integration of the 2PI dynamic equation is complicated by the required computational effort.
However, as we will show in the next chapter, this challenge can be met by use of semi-classical simulations of the field equations of motion.

Before we proceed, let us comment on the possibility to understand the turbulent scaling taking a renormalisation-group viewpoint~\shortcite{Berges:2008sr,Berges2012a,Philipp2012a}. 
Turbulence has served, since the seminal work of Kolmogorov~\shortcite{Kolmogorov1941a,Obukhov1941a,Frisch1995a}, as one of the first phenomena to develop renormalisation-group techniques out-of-equilibrium. 
The effectively local transport processes in momentum space, which are at the basis of turbulent cascades, immediately suggest themselves for a renormalisation group analysis. 
Building on functional renormalisation-group techniques \shortcite{Wetterich:1992yh,Berges:2000ew,Gies:2006wv,Pawlowski:2005xe} specifically out of equilibrium \shortcite{Canet:2003yu,%
Kehrein2004a,%
Mitra2006a,%
Zanella:2006am,%
Canet:2006xu,%
Gezzi2007a,%
Jakobs2007a,%
Korb2007a,%
Matarrese:2007wc,%
Karrasch2008a,%
Jakobs2009a,%
Schoeller2009a,%
Gasenzer:2008zz,%
Berges:2008sr,%
Gasenzer:2010rq}, 
more refined scaling analyses are being developed 
\shortcite{Canet2010a,Mathey2010a}.
The scaling exponents given in \Eq{kappaIR} result as canonical exponents, from power counting of flow equations for $G$ and higher-order vertices.
Additional flow equations which, in particular, account for the interaction effects in the spectral functions, are expected to fix potentially relevant anomalous dimensions.
We remark that the vortex picture developed in the following leads to the possibility to conjecture that such anomalous scaling will become important in driven systems where interaction effects between finitely spaced defects become relevant.

\subsection{Vortices in a superfluid}
\label{sec:VorticesinaSF}

Near degeneracy, the strongly occupied low-momentum modes of an ideal Bose gas, $n_\mathbf{k}\gg 1$, constitute the Rayleigh-Jeans distribution $n_\mathrm{RJ}(k) = {2mT}/{k^2}$ of classical waves.\footnote{In the following we use units where $\hbar=k_{B}=1$.} 
One can show that also out of equilibrium, strongly occupied modes can be described by the dynamics of classical waves~\shortcite{Brewczyk2007a,Berges:2007ym,Blakie2008a,Polkovnikov2010a}. 
The state of the gas can be defined in terms of the Wigner quasi-probability distribution $W[\phi,\phi^{*}]$ for the complex field $\phi(\vector{x})$ and its conjugate momentum $\phi^*(\vector{x})$ at each point in space.
In the considered wave-classical limit, $W$ is positive definite. 
Since quantum effects arising from coupling to sparsely occupied modes are small, the dynamics of the Wigner function follows a classical Liouville equation. 
In this semi-classical limit, the so-called truncated Wigner approximation allows to follow the evolution \emph{exactly}, within computational errors, by evaluating many trajectories evolving according to the classical field equation
\begin{equation}  \label{eq:GPE}
   \i \del_t \phi(\mathbf{x},t)= \left[ -\frac{\nabla^2}{2m}+ V(\mathbf{x})+g|\phi(\mathbf{x},t)|^2 \right] \phi(\mathbf{x},t) \,.
\end{equation}
While \eq{GPE} is equivalent in form to the Gross-Pitaevskii equation (GPE) for the quantum field expectation value $\langle\Phi\rangle$ the statistical sampling procedure leads to a quasi-exact result for the full many-body evolution.
Correlation functions are obtained by averaging over many trajectories. 
The non-linear classical field equation \eq{GPE} has some interesting properties relevant for our purposes:
It can be mapped to an Euler-type hydrodynamic equation, implying the interpretation of the gas dynamics in terms of (superfluid) flow. 
Among the possible solutions of this equation those resembling eddy flow and shock waves lead to topologically nontrivial configurations. 
These include vortices, solitons, and related nonlinear stationary states if more than two field components couple to each other, see, e.g. \shortciteN{Pismen1999a}, \shortciteN{Pitaevskii2003a}. 

\paragraph{Hydrodynamic representation} The polar representation $\phi= \sqrt{n}$ $\exp\{i\varphi\}$
allows to express the particle current $\mathbf{j}=i(\phi^*\nabla \phi - \phi \nabla \phi^*)/(2m)=n\mathbf{v}$ in terms of the velocity field $\mathbf{v}=m^{-1}\nabla\varphi$ and the particle density $n=|\phi|^{2}$. 
With this, the GPE \eq{GPE} can be rewritten into the continuity and an effective Euler equation for a compressible inviscid (i.e. non-viscous) fluid with modified pressure $m^{-1}\tilde{\mu}$,
\begin{eqnarray}
  \partial_t n + \nabla \cdot (n\mathbf{v})
  &=&0\,,
  \\
  \partial_t \mathbf{v} + (\mathbf{v} \cdot \nabla) \mathbf{v} 
  &=& -m^{-1}\nabla \tilde{\mu} \, , 
  \quad 
  \tilde{\mu} 
  \equiv gn-\frac{1}{2\sqrt{n}}\triangle\sqrt{n}\,.
\label{eq:Euler}
\end{eqnarray}
As the velocity is a potential field it is irrotational wherever the density is non-vanishing and therefore the phase $\varphi$ of the field $\phi$ well defined.
Small excitations of $\phi$ which give rise to small density fluctuations and small velocities are described by the linearised version of the hydrodynamic equations \eq{Euler}.
At wave lengths larger than the healing length $\xi=[2mng]^{-1/2}$ these are collective sound-wave excitations.\\ 

\noindent\textbf{Solitons}\ \ are quasi-topological one-dimensional solutions of \Eq{GPE} which travel with a fixed velocity but are non-dispersive, i.e., stationary in shape, see, e.g., \shortciteN{Zakharov1972a}, \shortciteN{Kevrekidis2008a}. 
For positive coupling constant $g>0$, the solitons are 'dark', i.e., characterised by an exponentially localised density depression in the surrounding bulk matter. 
This and the corresponding phase shift in complex field are given by
\begin{equation}
\label{eq:SingleSoliton}
\phi_{\nu}(x,t)=\sqrt{n}\left[\gamma^{-1}\tanh\left(\frac{x-x_{s}(t)}{\sqrt{2}\gamma\xi}\right)+i\nu\right]\,,
\end{equation}
where $x_{s}(t)=x_{0}+\nu t$ is the position of the soliton at time $t$.
Depending on the depth of the density depression, the dark soliton is either called grey or, for maximum depression, black. 
$\gamma=1/\sqrt{1-\nu^{2}}$ is the `Lorentz factor' corresponding to the velocity $v$ of the grey soliton in units of the speed of sound, $\nu=v/c_{s}=|\phi_{\nu}(vt,t)|/\sqrt{n}$. 
Being related to the density minimum, $\nu$ measures the `greyness' of the soliton,  ranging between $0$ (black soliton, $|\phi_{\nu}(vt,t)|=\nu\sqrt{n}=0$) and $1$ (no soliton, $|\phi_{\nu}(vt,t)|=\sqrt{n}$).
Due to the interaction with sound, solitons can continuously vanish which means that they are not topologically stable. 
In $d>1$ dimensions, solitons decay into vortices~\shortcite{Anderson2001a,Brand2002b}. 
The energy to create a soliton on top of a uniform background is $ E_\mathrm{s}={\left(1-\nu^2 \right)^{3/2}4}n c_{s} /3$~\shortcite{Pitaevskii2003a}.
For small velocities $\nu \ll 1$, this gives $E_\mathrm{s}\simeq 4 n c_{s}/3-2 n c_{s} \nu^2$ reminiscent of a classical point particle with negative mass $-4n/c_{s}$. 
The energy of a soliton monotonously decreases with increasing velocity which hints at a dynamical instability.\\

%
\noindent\textbf{Vortices}\ \ are topologically stable solutions of \Eq{GPE} in $d>1$ dimensions which form the superfluid analogies of eddy flows in classical fluids.
We recall that the ground-state manifold given by the minimum of the effective potential $V(\phi)=\mu|\phi|^2 + g|\phi|^4$ of the classical field  $\phi(\vector{x}) = \sqrt{n(\vector{x})}\exp\{i\varphi(\vector{x})\}$ requires constant density $n(\vector{x})=n$, but is degenerate in the phase $\varphi(\vector{x})$. 
The true ground state has a constant phase, and is therefore called topologically trivial. 
On the other hand, we can consider field configurations which have constant density on the boundary of, e.g., a two dimensional volume, but a varying phase. 
If we use this freedom to evolve the phase angle $\varphi(\vector{x})$ from $0$ to $2\pi$ when going around the boundary, we arrive at a topologically nontrivial state. 
Configurations are topologically distinct, because one can not define a continuous function that transforms one into the other.
As a consequence of the phase winding  the phase can not be well-defined at some point inside the volume, and hence, the density has to go to zero at that point. 
The stationary state of \Eq{GPE} which exhibits these properties is called a vortex~\shortcite{Pitaevskii1961a}.
Following the phase angle along a closed path around the vortex core it continuously varies from $0$ up to $\varphi=2\pi\kappa$, where the integer $\kappa$ is called the winding number or circulation. 
Only a singly quantised vortex with $\kappa= \pm 1 $ is stable. 
It is described, in polar coordinates centred at the vortex core, by the field $\phi(r,\varphi)=f(r)\exp\{i\kappa\varphi\}$ where $f(r)$ can be chosen real and approaches the square root of the bulk density $n$ for large distances $r$ from the vortex core.
At $r \rightarrow 0$, $f(r) \sim r$ rises linearly. 
$\phi(r,\varphi)$ is a stationary solution of \Eq{GPE}, evolving as $\phi(\mathbf{r},t)=\phi(\mathbf{r}, 0)\exp\{-i\mu t\}$ with  $\mu=gn$. 

We remark that the irrotational nature of the velocity field defined in the hydrodynamic formulation of the GPE is restricted to those points where the density is non-vanishing.
The velocity field of the vortex is, in the polar coordinates used before, $ \tilde{\mathbf{v}}(r,\varphi) = \kappa \mathbf{e}_{\varphi}/(mr) $.
Its curl is concentrated locally at the vortex core.
The compact phase $\varphi \in [0,2\pi)$, becomes a non-compact velocity potential. 
As a consequence, vortex creation or annihilation is not described by the effective Euler equation. 
In fact, due to the Thomson circulation theorem vorticity is locally conserved in an inviscid flow~\shortcite{Lesieur2008a}.

The energy associated to the vortex is extremely non-local:
For a singly quantised vortex in a 2D homogeneous gas the energy within the volume $V=\pi R^2$ grows logarithmically with the radius,  $E_\mathrm{v}={\pi n}{m^{-1}} \mathrm{ln}\left(1.46R/\xi\right)$, see, e.g., \shortciteN{Pitaevskii2003a}.
In $d=3$ dimensions, point vortices extend to vortex lines around which the fluid rotates \shortcite{Pitaevskii1961a,Pismen1999a}. In the simplest case, a vortex line of length $L$ goes straight from one end of the volume to the other. This requires the excitation energy $LE_\mathrm{v}$. Vortex lines can not end inside the medium, but can form closed loops of all shapes, e.g. rings, ellipses, also knots. 
The GPE \eq{GPE} moreover supports linear wave excitations of the position of the vortex lines, so called Kelvin waves~\shortcite{Sonin1987a,Kivotides2001a,Vinen2003a,Krstulovic2012a}. 
Higher dimensional vortices exist whereby the dimensionality of vortex-core geometry is always $d-2$, for example leading to vortex surfaces in four dimensions.

\paragraph{Decomposition of the flow field}
A non-equilibrium flow features the presence of multiple types of excitations. In order to distinguish longitudinal excitations (sound waves) from rotational excitations (vortices), we close this section by discussing a decomposition of the kinetic energy density proposed by \shortciteN{Nore1997a}. 
The total kinetic energy $E_{\mathrm{kin}}= {m} \int \mathrm{d}^dx \, \langle |\nabla \phi(\mathbf{x},t)|^2\rangle/2$ can be split, $E_{\mathrm{kin}} = E_{\mathrm{v}} + E_\mathrm{q}$, into a `classical' part $E_\mathrm{v}= {m}\int \mathrm{d}^dx \, \langle |\sqrt{n}\mathbf{v}|^2 \rangle/2 $ and a `quantum-pressure' component $E_\mathrm{q}=\int \mathrm{d}^dx \, \langle |\nabla \sqrt{n}|^2 \rangle/(2m) $.
The radial energy spectra for these fractions involve the Fourier transform of the generalised velocities $\vector{w}_{\mathrm{v}}=\sqrt{n}\vector{v}$ and $\vector{w}_{\mathrm{q}}=\nabla\sqrt{n}/m$,
\begin{eqnarray}
 E_{\delta}(k)= \frac{m}{2} \int \mathrm{d}\Omega_d \, \langle |\mathbf{w}_{\delta}(\mathbf{k})|^2 \rangle,\quad \delta=\mathrm{v},\mathrm{q}.
\end{eqnarray}
which we cast further into occupation numbers $n_{\delta}(k)=k^{-2}E_{\delta}(k), \delta=v, q$. 
Since the superfluid velocity $\mathbf{v}=\nabla\varphi$ is a potential field it does not reveal a transversal flow component, $\nabla \times \tilde{\vector{v}} = 0$ (outside vortex cores). 
On the contrary, $\mathbf{w}_{\mathrm{v}}$ is not a potential field and the divergence of $\mathbf{v}$ at $r \rightarrow 0$ is regularised by the vanishing $\sqrt{n(r)}$. 
Following \shortciteN{Nore1997a} the regularised velocity $\mathbf{w}_{\mathrm{v}}$ can be further decomposed into `incompressible' (divergence-free) and `compressible' (solenoidal) parts, $\mathbf{w}_{\mathrm{v}}=\mathbf{w}_{\mathrm{i}}+\mathbf{w}_{\mathrm{c}}$, with $\nabla \cdot \mathbf{w}_\mathrm{i}=0$, $\nabla \times \mathbf{w}_\mathrm{c}=0$, to distinguish vortical superfluid and rotationless motion of the fluid. 
By construction, the generalised velocity $\tilde{\mathbf{w}}_{\mathrm{v}}=f(r) \tilde{\mathbf{v}}(r,\varphi)$ of a vortex 
has only an incompressible component, since
\begin{eqnarray}
\nabla \cdot \tilde{\mathbf{w}}_{\mathrm{v}} =  \nabla f(r) \cdot \tilde{\mathbf{v}}(r,\varphi) + f(r) \nabla \cdot \tilde{\mathbf{v}}(r,\varphi)=0 \,.
\end{eqnarray}
The first term vanishes due to the transversal nature of the vortex velocity field, the second one equals zero because the superfluid velocity is a potential field. 
The density of incompressible energy $|\tilde{\mathbf{w}}_{\mathrm{i}}|$ of a vortex is constant up to about one healing length distance from the core and then falls off as $1/r$.
Sound waves
are purely compressible excitations. 
In our simulations, their oscillating density and phase profiles will be visible as maxima and minima in the compressible energy density in position space.

\subsection{Vortex statistics}

To understand the implications of vortex defects appearing in the dynamical evolution of degenerate Bose gases, in particular to make contact to the observables studied in the context of wave turbulence we turn to a statistical viewpoint. 
The point vortex model studied in the following was introduced by \shortciteN{Onsager1949a}.
It describes the complex flow pattern in terms of the statistics of classical point objects with Coulomb-type interactions.
However, due to the absence of a kinetic energy term in the Hamiltonian there is no kinematic transfer of potential into motional energy. 
The model is constructed as a discrete-vorticity approximation of classical fluid turbulence, but it is even more suitable to describe superfluid turbulence consisting of quantised vortices. 

We restrict our discussion to the example of $d=2$ dimensions. 
We have seen that an isolated, singly quantised vortex is described by the complex field $\phi(r,\varphi) \equiv \sqrt{n(r)} e^{i\varphi}$.
As the $r$-dependence of the density $n(r)$ only becomes important at small scales on the order of the healing length $\xi=(2mgn)^{-1/2}$ at which in practice thermal excitations dominate, we assume $n$ to be uniform.
A set of $M$ vortices can be described by $\phi(\mathbf{x}) = \Pi_i^M \phi_i(\mathbf{x})$, where $\phi_i(\mathbf{x}) = \phi(\mathbf{x}-\mathbf{x}_i)$ is the single-vortex field centred around $\mathbf{x}_i$. 
Let us derive the corresponding bosonic single-particle spectrum by considering the velocity field $\mathbf{v} =  \nabla \varphi/m$. 
We can express the mean classical kinetic energy density of the velocity field $\tilde{\mathbf{v}}(\mathbf{x})$ of a single vortex as 
\begin{eqnarray}  \label{eq:EnergySpec}
E_\mathrm{v}(\mathbf{x}) 
&=& \frac{m}{2}  \, \langle |\,  \mathbf{v}(\mathbf{x}) |^2 \rangle 
= \frac{m}{2} \, \langle |\, \int \mathrm{d}^2x' \,  \tilde{\mathbf{v}}(\mathbf{x} - \mathbf{x'}) \, \rho(\mathbf{x'}) \, \, |^2 \rangle \, ,
\end{eqnarray}
where $\rho(\mathbf{x}) = \sum_{i=1}^{M} \kappa_i \delta(\mathbf{x}-\mathbf{x}_i)$ defines the spatial distribution of vortices with winding number $\kappa_i = \pm 1$. 
Here and in the following, $\langle\cdot\rangle$ denotes an ensemble average over different realisations of the classical field $\phi(\mathbf{x})$.
We derive the low-$k$ scaling of $n(k)$ from the kinetic-energy spectrum $E_\mathbf{v}(k)$, the angle-averaged Fourier transform of $E_\mathbf{v}(\mathbf{x})$,  taking into account that at low $k$,  the velocity field $\mathbf{v}$ dominates the dynamics,  
\begin{equation}  \label{eq:nkfromEvk}
  n(k) \simeq 2mk^{-2} E_\mathbf{v}(k).
\end{equation}
One has, from  \Eq{EnergySpec},
\begin{equation}  \label{eq:Evk}
 E_\mathrm{v}(\mathbf{k}) \sim \langle |\mathbf{v}(\mathbf{k})|^2 \rangle = \langle  \, |\rho(\mathbf{k})|^2  \, |\tilde{\mathbf{v}}(\mathbf{k})|^2 \rangle \,,
\end{equation}
with 
\begin{equation}  \label{eq:Vortexdensity}
|\rho(\mathbf{k})|^2  =  \sum_{i,j}^M  \kappa_i \kappa_j e^{i\mathbf{k}(\mathbf{x}_i-\mathbf{x}_j)} \,.
\end{equation}
Below the healing length scale $k_{\xi} = 2\pi/\xi$,
the modulus of the velocity field of a single vortex scales as $ | \tilde{\mathbf{v}} | \sim k^{-1} $ and is radially symmetric.

To distinguish contributions from vortex-vortex and vortex-antivortex correlations we write the distribution $\rho(\mathbf{x}) = \rho^\v(\mathbf{x}) - \rho^\a(\mathbf{x})$ as the sum of distributions $\rho^{\v}(\mathbf{x})= \sum_{i=1}^{M} \delta( \mathbf{x}-\mathbf{x}_i^\v)$ of $M$ vortices and $\rho^\a(\mathbf{x})= \sum_{i=1}^{M} \delta(\mathbf{x}-\mathbf{x}_i^\a)$ of $M$ antivortices. 
Hence,
\begin{eqnarray}  \label{eq:CF}
\langle \, |\rho(\mathbf{k})|^2 \, \rangle  = \int \mathrm{d}^2x \, \mathrm{d}^2x' \, e^{i\mathbf{k}(\mathbf{x}-\mathbf{x'})} C(\mathbf{x},\mathbf{x}') \,,
\end{eqnarray}
with $C(\mathbf{x},\mathbf{x}')= \langle \, \rho^\v_\mathbf{x} \rho^\v_\mathbf{x'} \, \rangle - \langle \,\rho^\v_\mathbf{x} \rho^\a_\mathbf{x'}  \, \rangle- \langle \,\rho^\a_\mathbf{x} \rho^\v_\mathbf{x'}  \, \rangle + \langle \,\rho^\a_\mathbf{x} \rho^\a_\mathbf{x'} \, \rangle$.
This allows for a derivation of the kinetic-energy distribution in terms of correlation functions of vortex positions. 

Now, we can model pairing by the density-density correlation functions
\begin{eqnarray}
\langle  \rho^{\v(\a)}_{\mathbf{x}} \rho^{\v(\a)}_{\mathbf{x'}} \rangle 
&=& \frac{M}{V_R}\delta(\mathbf{x}-\mathbf{x'}) + P_{\mathbf{x},\mathbf{x}'},  
\label{eq:rhoVrhoV}
\\
\langle  \rho^{\v(\a)}_{\mathbf{x}} \rho^{\a(\v)}_{\mathbf{x'}} \rangle 
&=& \frac{M}{V_RV_\lambda}\theta(\lambda - |\mathbf{x}-\mathbf{x'}| ) +  P_{\mathbf{x},\mathbf{x}'}  
\label{eq:rhoVrhoA}
\end{eqnarray}
where $V_{R}$ is the volume in which we take averages, and $V_\lambda= \pi \lambda^2$ is the area where the theta function equals one, measuring the correlation regime of vortices and antivortices.  
The contributions
\begin{eqnarray}
P_{\mathbf{x},\mathbf{x}'} &=&  \frac{M\left({M}-1 \right)}{V_{R}(V_{R}-V_{\Lambda})}\theta(|\mathbf{x}-\mathbf{x'}| -\Lambda) 
\end{eqnarray}
take into account that, besides pairing, vortices and antivortices keep a minimum distance $\Lambda$ in the dilute gas. 
This is due to vortex-vortex repulsion and fast vortex-antivortex annihilation on small distances. 
The functions $P_{\mathbf{x},\mathbf{x}'}$ cancel out in \Eq{CF}.\footnote{If different avoidance scales $\Lambda$ apply for vortices and antivortices, the terms do not cancel, but the remaining term does not alter the results for pair scaling derived here.}

From this ansatz, two scaling regimes can be found \shortcite{Nowak:2011sk}. 
In the case of pairing, the flow field far away from the cores is given by the field of a vortex pair which decays as $r^{-2}$, and the low-momentum power law is dominated by the flow of random vortex pairs $ n_k \sim k^{-2} $. 
Above $k_{\mathrm{pair}}\simeq \pi/\lambda $, the distribution exhibits the scaling of an ensemble of independent vortices, $n_k \sim k^{-4}$, up to the healing-length scale $k_\xi$ above which one can observe the vortex-core scaling $\sim k^{-6}$.
The above results show that, in a vortex dominated flow, particles with low momenta are found far away from the vortex cores. 
Particles closer to the vortex cores pick up a higher momentum. 

We remark for conciseness, as was shown by \shortciteN{Novikov1976a}, that one can obtain Kolmogorov 5/3 scaling from the statistics of point vortices, by choosing the density-density correlation functions to decay as
\begin{eqnarray}
\langle  \rho^{\v(\a)}_{\mathbf{x}} \rho^{\v(\a)}_{\mathbf{x'}} \rangle &\sim& |\mathbf{x}-\mathbf{x'}|^{-\alpha} + P_{\mathbf{x},\mathbf{x}'}\, ,  
\label{eq:K41rhoVrhoV}
\\
\langle  \rho^{\v(\a)}_{\mathbf{x}} \rho^{\a(\v)}_{\mathbf{x'}} \rangle &=& P_{\mathbf{x},\mathbf{x}'}\,
\label{eq:K41rhoVrhoA}
\end{eqnarray}
where the contributions $P_{\mathbf{x},\mathbf{x}'}$ are assumed to be equal. 
The integral in \Eq{CF} is convergent for $1/2 <\alpha <2$. 
This includes $\alpha=4/3$ which gives $n(k)\sim k^{-4.66}$ and  thus Kolmogorov scaling $E(k)\sim k^{-5/3}$. 
Note that the presence of vortex-antivortex correlations destroys the $5/3$ scaling as discussed by \shortciteN{Bradley2012a}.

\paragraph{Infrared cutoff}
%
 \begin{figure}
 \center
 \includegraphics[width=0.6\textwidth]{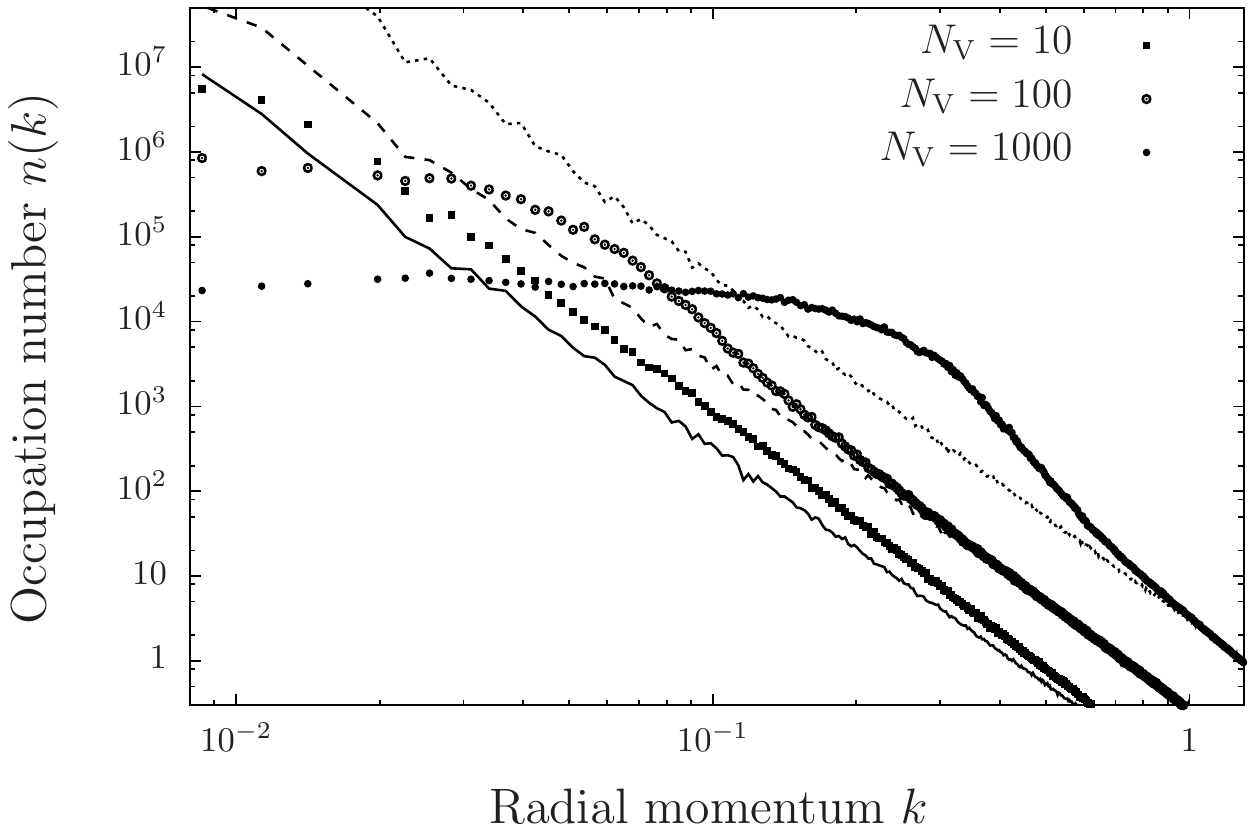}
 \caption{Radial momentum distributions for a set of infinitely thin ($\xi=0$) randomly distributed vortices. 
 Note the log-log scale. 
 At the highest momenta, the total number spectrum $n(k)$ (dots) and the incompressible component $n_i(k)$ (lines) follow the $k^{-4}$ scaling of independent vortices. 
 However, at momenta $k<k_\mathrm{V}$, with $k_\mathrm{V}(N_\mathrm{V}=10) \simeq 10^{-2}$, $k_\mathrm{V}(N_\mathrm{V}=100)$ $ \simeq 3 \times 10^{-2}$, $k_\mathrm{V}(N_\mathrm{V}=1000) \simeq 10^{-1}$,  $n(k)$ becomes flat while $n_i(k)$ continues to rise.
}
 \label{fig:IR-cutoff}
\end{figure}
%
onsider a Bose gas with density $n_\mathrm{V}$ containing a random distribution of vortices of either sign. 
We expect a decay of the coherence $\langle \phi^{*}(\mathbf{x}) \phi(\mathbf{x'})\rangle$  
over a distance $|\mathbf{x}-\mathbf{x'}|$ of the order of the mean vortex distance $l_\mathrm{V} = n^{-1/2}_\mathrm{V}$, corresponding to a momentum scale $k_\mathrm{V}=\pi/l_\mathrm{V}$.  
That is because vortices appear on average at this distance and induce a rapid change of the phase angle $\varphi$. 
For momenta $k<k_\mathrm{V}$, the momentum distribution $n(k)$ needs to be sufficiently flat in order to insure convergence of the integral that gives the total number of particles. 
This restriction is not imposed upon the particle numbers defined by the hydrodynamic decomposition in terms of $n_i, n_c, n_q$. 
Hence, we expect a deviation of $n(k)$ from the incompressible momentum distribution $n_i$ below $k_\mathrm{V}$.
In \Fig{IR-cutoff}, we present numerical evidence for our reasoning. The plot shows the single-particle (dots) as well as the incompressible momentum distributions (lines) for three different vortex numbers $N_\mathrm{V}$. 
One can observe the vortex density dependence of the IR cutoff in the single-particle momentum distribution. 
The spectrum of the incompressible velocity field does not show this feature. 
Instead, the $k^{-4}$ scaling persists all the way to the lowest momenta.

\paragraph{Onsager model: Non-thermal fixed point as a maximum-entropy state}
%
 \begin{figure}
 \center
 \includegraphics[width=0.8\textwidth]{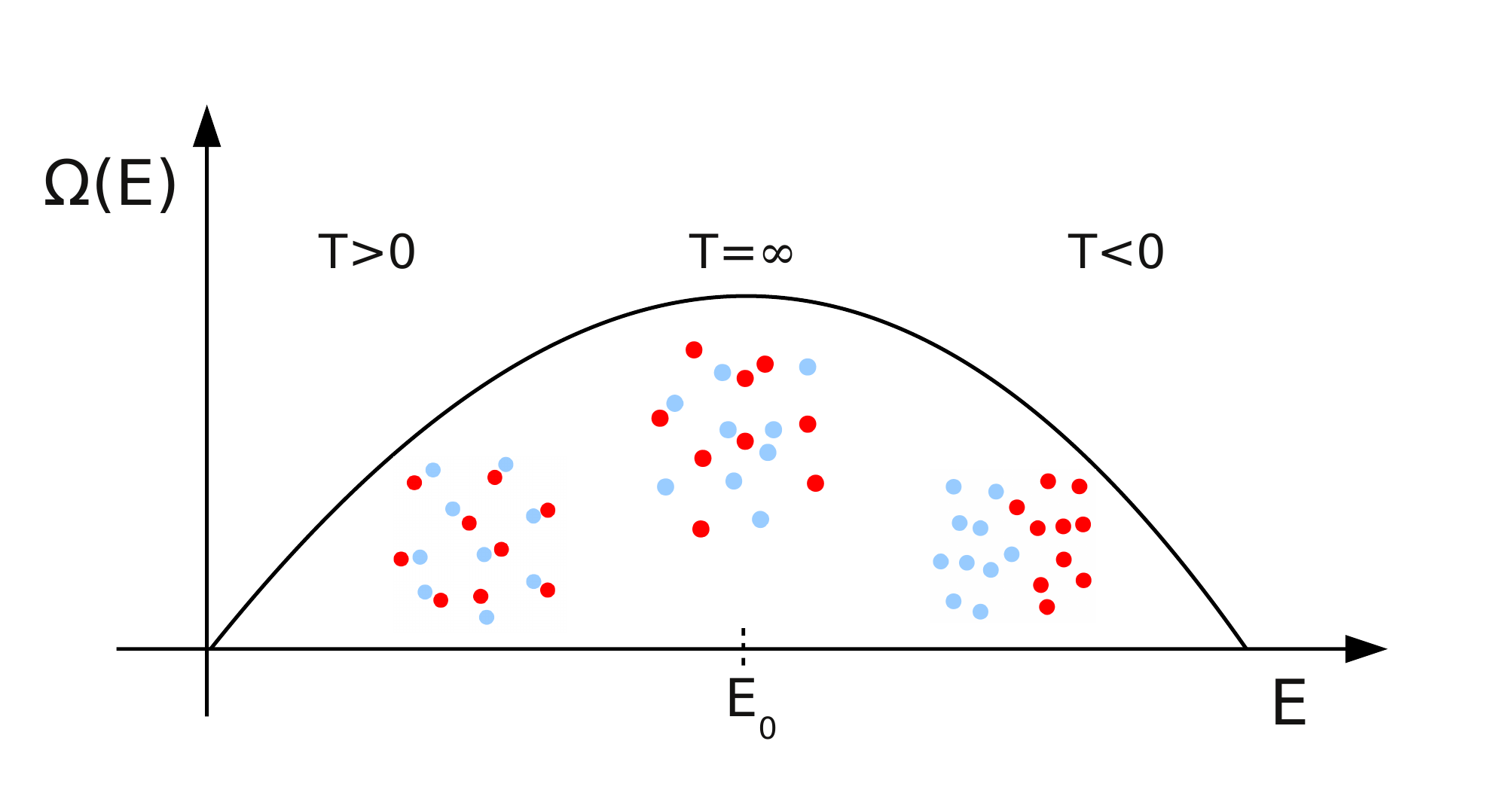}
 \caption{Illustration of Onsager's picture of thermodynamic equilibrium states of the balanced two-dimensional Coulomb-type vortex gas at fixed defect number. 
 We plot the qualitative form of the number of available states $\Omega(E)$ as a function of energy $E$ as well as snapshots of typical microscopic configurations of vortices and antivortices.
}
 \label{fig:Onsager}
\end{figure}
%

We close with the picture \shortciteN{Onsager1949a} developed of thermodynamic equilibrium states of a fixed number of vortices and antivortices in two dimensions.
He used the Hamiltonian of vortical flow \shortcite{Lin1941a},
\begin{eqnarray}  \label{eq:Hamiltonian}
H= -\frac{1}{2\pi} \sum_{i>j}^M \kappa_i \kappa_j \mathrm{ln}(|\mathbf{r}_{i} -\mathbf{r}_{j}|) \,,
\end{eqnarray}
to describe the dynamics of a system of $M$ vortices in a superfluid which hence interact like charge carriers in a Coulomb gas.
Here, the position of the $i$-th vortex is denoted as $\mathbf{r}_{i}=(x_i,y_i)$. 
Due to the fact that the $x$ and $y$ coordinates of each vortex are canonical conjugates of each other, phase space is identical with position space. 
Hence, for $M$ vortices moving in a  2D volume $V$ the total phase space is $V^M$. 
The Hamiltonian \eq{Hamiltonian} implies that low-energy configurations feature vortices of opposite sign close to each other, whereas high-energy configurations require vortices of equal sign to group. 
Due to these constraints, the number of configurations $\Omega(E)$ available for the system at a given energy $E$ decreases towards high and low energies, with a maximum at some intermediate $E=E_0$. This concept is illustrated in \Fig{Onsager}.
According to Boltzmann, the entropy is 
\begin{equation}
 S(E)= \mathrm{ln}\left[\Omega(E)\right]\,,
\end{equation}
and the inverse temperature $1/T=\del S/\del E$ is positive for $E<E_0$ and negative for $E>E_0$. 
It follows that positive-temperature states are characterised by vortex-antivortex pairing, while negative-temperature states feature vortices of the same circulation to cluster. 
At the point of maximum entropy $S(E_0)$ and infinite temperature, Onsager expected a state of uncorrelated vortices and antivortices. 

\paragraph{3-dimensional systems}

We briefly comment on the case of vortex lines and loops in three dimensions.
A formulation similar to the Onsager point vortex model is possible~\shortcite{Nemirovskii1998a,Nemirovskii2002a,Tsubota2008a}. 

For taking into account the most general case of squeezed vortex loops, it is helpful to consider elliptical filaments, characterised by  a major radius $r_a$ and minor radius $r_b$. 
Three scaling regimes can be distinguished. 
For the lowest momenta, one has $n(k) \sim k^{-2}$, which equals the infrared scaling in the presence of a vortex ring. 
For momenta  $k_{a} \ll k \ll k_{b}$, one finds $n(k) \sim k^{-3}$, which coincides with the infrared scaling of two anti-circulating vortex lines. 
For the ellipse, above $k_{b}$, the momentum distribution scales like $n(k) \sim k^{-5}$.  
This is the scaling of a single vortex line and can also be found as the high-momentum scaling of a vortex ring or a pair of straight vortex lines.
For more details we refer to \shortcite{Nowak:2011sk}.

\section{Non-thermal fixed point of a vortex gas}

In extending the concept of universality to time evolution far from thermal equilibrium one expects that also away from the thermal limit the character of dynamical evolution can become independent of the microscopic details. 
Looking at closed systems this implies that, in approaching critical configurations, the evolution must become independent of the particular initial state the system has started from and critical slowing down in the actual time evolution is observed.  
Time evolution near the fixed point becomes equivalent to a coarsening transformation. 
It should look like a self-similar pattern which is observed through a microscope while one continuously turns the magnification of the lens to smaller focal lengths. 
Precisely at the fixed point the system becomes stationary due to its self-similarity under time translations. 
Considering a generic isolated system, it can evolve into the vicinity of a non-thermal fixed point, stay there over long times before it eventually undergoes thermalisation.
This picture is supported by the derivation of the strong turbulence scaling laws in the frame of renormalisation-group theory for correlation functions \shortcite{Berges:2008sr}.

In this section we discuss mostly numerical results obtained for the time evolution of an isolated two-dimensional gas.
This is characterized, from the turbulence point of view, by direct and inverse cascades, with fluxes determined by local conservation laws in Fourier space.
From the perspective of defect formation a diluting ensemble of vortices and antivortices marks the approach of the non-thermal fixed point.
We identify the mechanism for this dilution process and demonstrate that it is consistent with the coarsening transformation picture of universal dynamics.
By reducing the characterization of the momentaneous configuration to a few length parameters we can illustrate the slowing of the evolution near the fixed point.
We show that it depends on the chosen initial state how closely the critical point is approached.

\subsection{Time evolution of vortex patterns and momentum spectra}

\begin{figure}[!t]
\center
 \includegraphics[width=0.7\textwidth,height=0.44\textwidth]{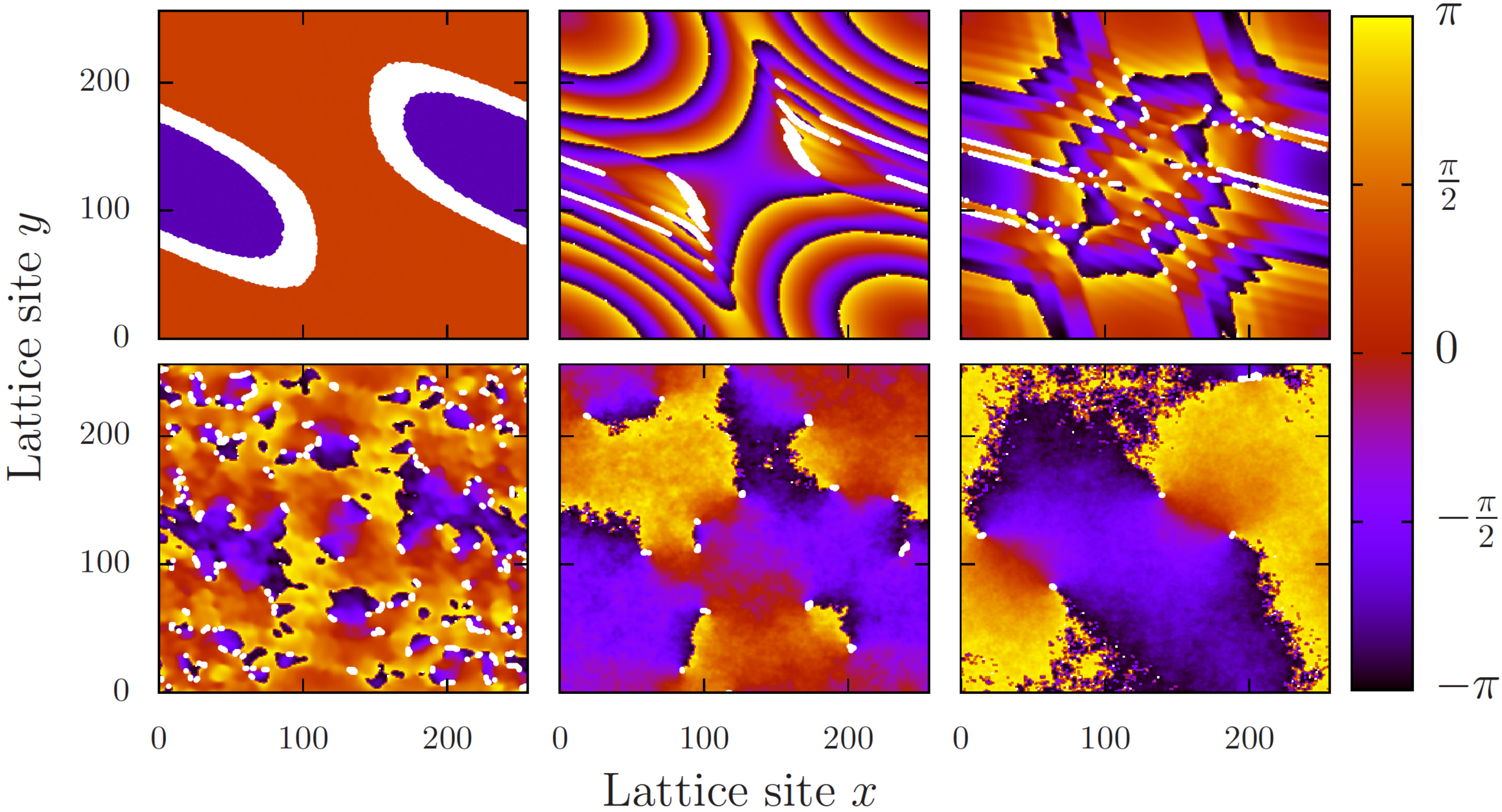}\\
\vspace{-0.03\textwidth}\hspace{0.022\textwidth}
 \includegraphics[width=0.73\textwidth,height=0.445\textwidth]{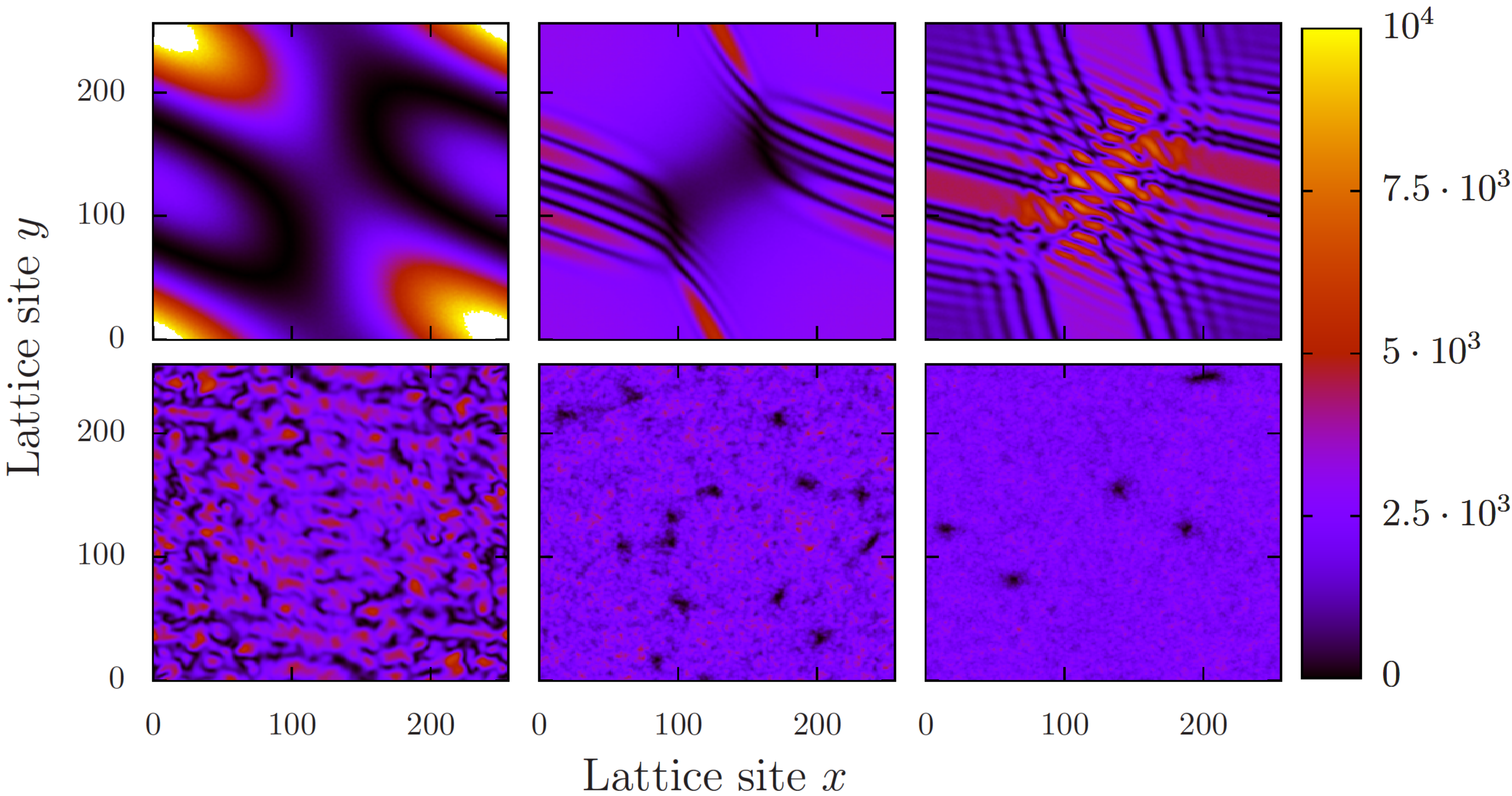}
 \caption{Phase angle $\varphi(\vector{x},t)$ (upper panel) and spatial density $n(\mathbf{x})$ (lower panel) at six times during a single run of the simulations in $d=2$ on a space-time lattice with side lengths $L=N_s a_s$, lattice spacing $a_s$, imposing periodic boundary conditions. 
The GPE is written in terms of the dimensionless variables:  $\overline{g}=2mga^{2-d}_s$, $\overline{t}= t/(2ma^2_s)$ and $\overline{\phi}(\overline{t})=\phi \sqrt{a^d_s} \mathrm{exp}(2i\overline{t})$. 
 Parameters are: $\overline{g}=3\cdot 10^{-5}$, $N=10^8$, $N_s=256$. 
Shown times are (by line, each from left to right) : 
$\overline{t}=0$: Ordered phase at initial preparation. 
$\overline{t}=104$: During build-up of strong density and phase gradients. 
$\overline{t}=210$: Formation of vortices and antivortices. 
$\overline{t}=420$: Unbinding of vortex-antivortex pairs.
$\overline{t}=6550$: Slowing down of vortex dynamics near the non-thermal fixed point.
$\overline{t}=10^5$: Few seperated vortex antivortex pairs close to non-thermal fixed point.}
\label{fig:2dPhase}
\end{figure}
%
 
%
 \begin{figure}[!t]
 \center
 \includegraphics[width=0.7\textwidth]{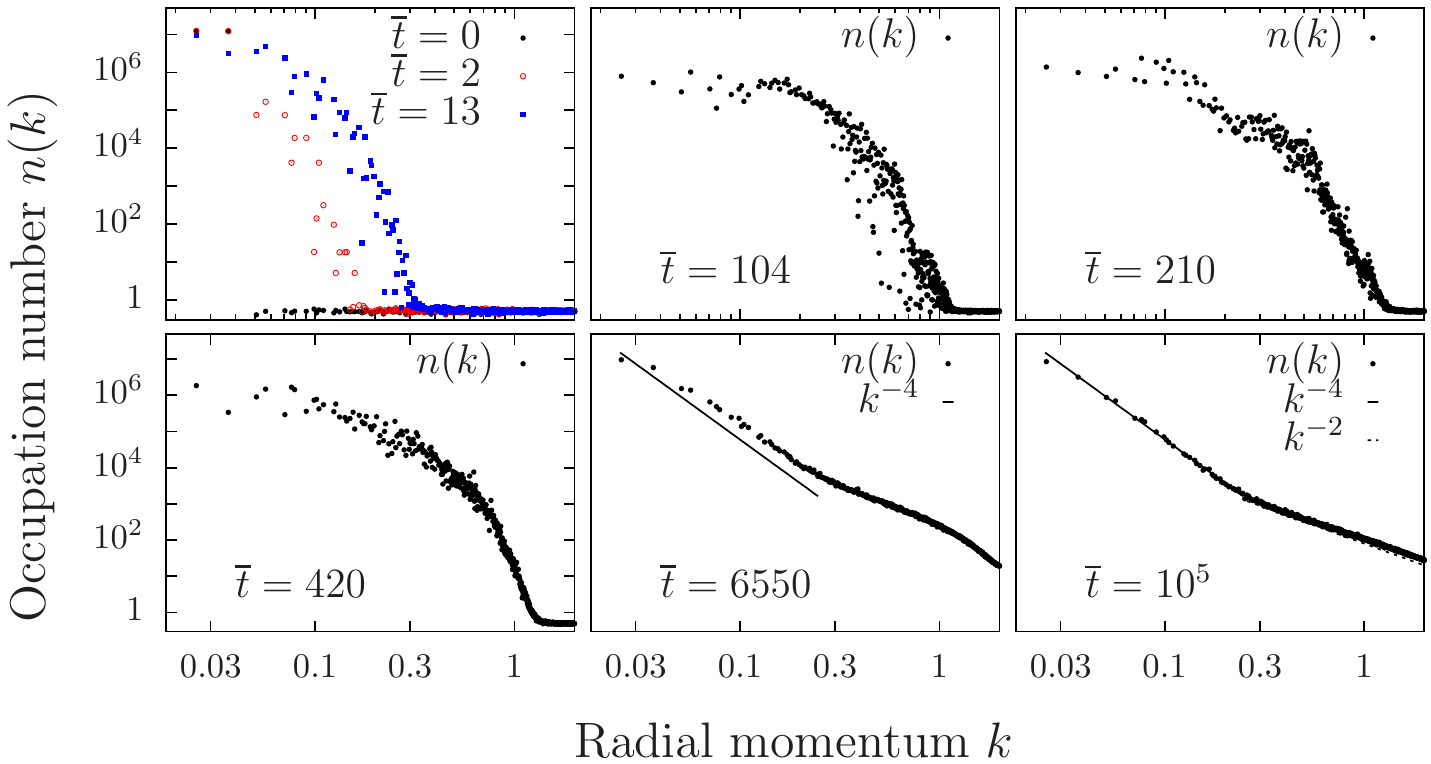}
 \caption{Single-particle, angular and ensemble averaged mode occupation numbers 
 $n(k) = \int \mathrm{d}^{d-1}\Omega_{k} \, \langle \phi^*(\mathbf{k})\phi(\mathbf{k}) \rangle_{\mathrm{ensemble}}$
 as functions of the radial momentum $k$, at the same evolution times as in  \Fig{2dPhase}. 
 The dimensionless lattice momenta are $k=[\sum_{i=1}^d 4\mathrm{sin}^2 (k_i/2)]^{1/2}, {k}_{i}=2\pi{n}_{i}/N_s, \,  n_i= -N_s/2,...,N_s/2$.
Parameters are the same as in \Fig{2dPhase}. 
Note the double-logarithmic scale. 
At early times (top left), scattering between macroscopically occupied modes leads to excitation transfer to high momenta. 
Vortex formation ($\overline{t}\simeq 210 $) sets in when this process has reached the healing length scale $k_\xi\simeq{1.26}$. 
At  $\overline{t}\gtrsim 10^4$ a bimodal power law emerges, characteristic for  the non-thermal fixed point.}
\label{fig:ModeOccupationSequenceEarly}
 \end{figure}
%
%
 %
In this section we have a closer look at the process of vortex formation and of the Bose gas approaching the non-thermal fixed point, concentrating again on the two-dimensional case.
Vortical excitations can be created in large numbers, e.g., within shock waves forming during the non-linear evolution of a coherent matter-wave field. 
We follow the exemplary evolution of phase and density profiles in \Fig{2dPhase}. 
Six snapshots are shown, taken at the dimensionless times $\bar t$ as indicated in the caption. 
The initial field configurations were prepared by macroscopically populating a few of the lowest momentum modes in the computation such that the resulting condensate density in position space varied between zero and some maximum value. 
One observes strong phase gradients forming due to the non-linear evolution. 
At around $t\simeq 100$, these gradients produce shock fronts delimited by phase defects which in the following collapse into vortex trains. 
Scattering processes between vortices quickly isotropise phase and density fluctuations. 

In \Fig{ModeOccupationSequenceEarly}, we show the corresponding time evolution of the angle- and ensemble-averaged radial momentum spectrum. 
The early times (top left) are characterised by scattering between macroscopically occupied modes. 
Once excitations of the order of the healing length $\xi$ are created, vortex formation sets in.
Shortly after vortices are created the spectrum exhibits a power-law behaviour $2.85 \lesssim \zeta \lesssim 3.0$ within a range of momenta $k \in [0.04:0.4]$, see the lower left panel of \Fig{ModeOccupationSequenceEarly}.  
Subsequently, the evolution slows down and a quasi-stationary period is entered. 
During an intermediate stage (bottom centre and right panels of  \Fig{ModeOccupationSequenceEarly}) of the vortex-bearing phase two distinct power laws develop which are in excellent agreement with the analytical prediction in Eqs.~\eq{kappaUV} and \eq{kappaIR}. 
While in the ultraviolet the exponent $\zeta_{P}^{\mathrm{UV}} = d$ exhibits weak wave turbulence, \Eq{kappaUV}, in the infrared, the exponent confirms the field theory prediction $\zeta_{Q}^{\mathrm{IR}}=d+2$, cf.~\Eq{kappaIR}. 
%
During the ensuing evolution, the weak-wave-turbulence scaling decays towards $\zeta=2$, reflecting a thermal UV tail.  
Note that in $d=2$, the weak-turbulence exponent $\zeta_{P}^{\mathrm{UV}}=2$ is identical to that in thermal equilibrium in the Rayleigh-Jeans regime, $n(k) \sim T/k^2$ \shortcite{Zakharov1992a}. 
In $d=3$ we observe, at late times, a change of the infrared scaling behaviour from $\zeta=d+2=5$ to $\zeta=3$, pointing to the development of pairing correlations \shortcite{Nowak:2011sk}.

At late times, after the last vortical excitations have disappeared, we observe the entire spectrum to become thermal, i.e., exhibit Rayleigh-Jeans scaling with $\zeta=2$ (not shown).
We emphasise that thermal scaling of the single-particle occupation number has $\zeta=2$ despite the fact that quasi-particles with a linear dispersion are expected to thermalise in the regime of wave numbers smaller than the inverse healing length.

\subsection{Local transport in momentum space and inverse particle cascade}
\label{sec:Fluxes}
%

 \begin{figure}[!t]
 \center
 \includegraphics[width=0.45\textwidth]{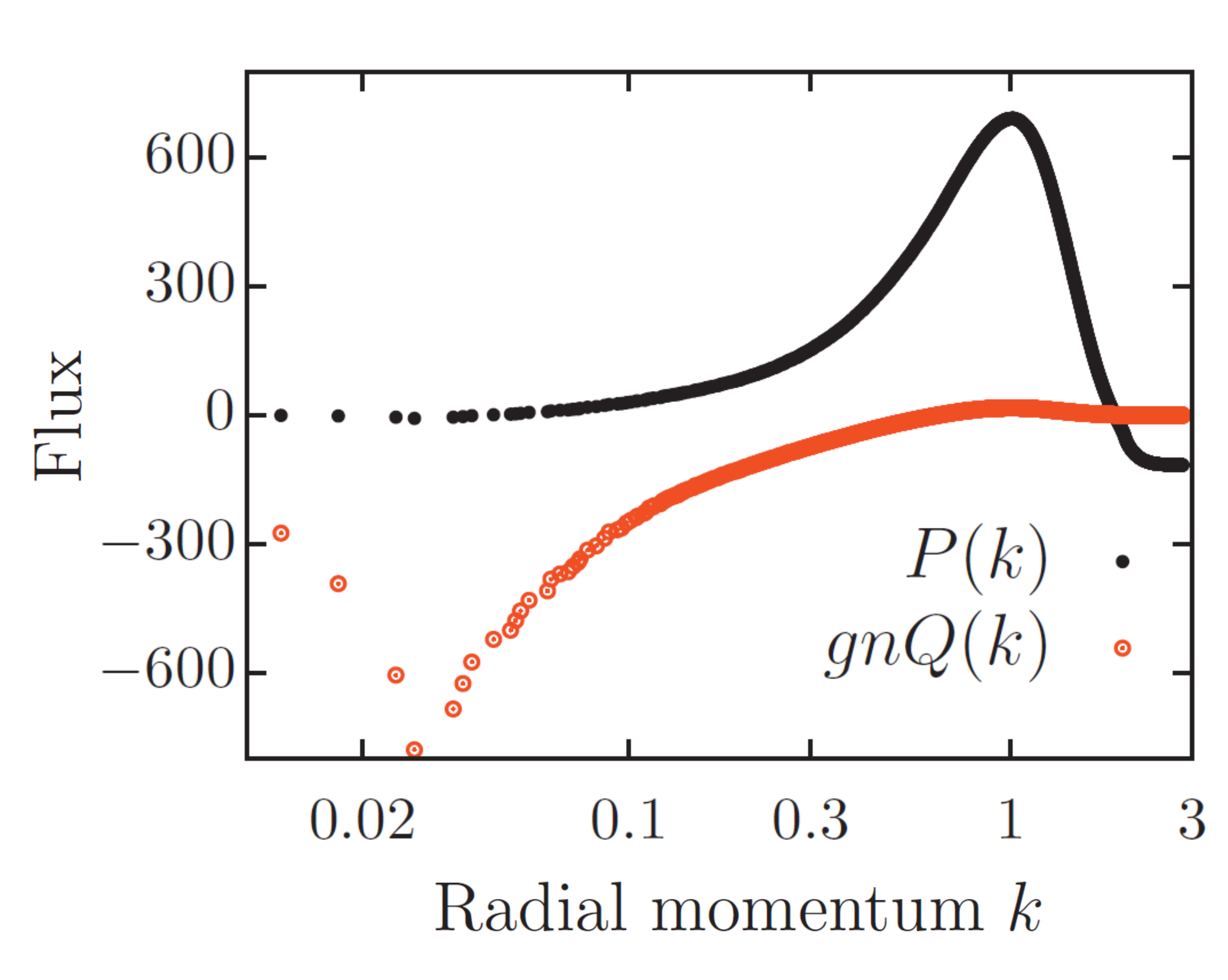}
 \caption{ Direct kinetic-energy and inverse particle fluxes in $d=2$ at $\overline{t}=6550$ {\protect\shortcite{Nowak:2011sk}}. 
 Parameters are as in \Fig{ModeOccupationSequenceEarly}. 
 Note the logarithmic $k$-axis. 
 A positive kinetic energy flux is seen in the UV, a negative particle flux in the IR. 
 Units: $[P]=[gnQ]=(4m^2a_s^{d+4})^{-1}$.
}
 \label{fig:PFlux2D3D}
 \end{figure}
%

From the above findings the question arises, why the system selects the particular exponents $\zeta_{P}^{\mathrm{UV}} = d$ and $\zeta_{Q}^{\mathrm{IR}}=d+2$ from the set of four possible exponents given in Eqs.~\eq{kappaUV} and \eq{kappaIR}.
For this, the fluxes underlying the stationary but non-equilibrium distributions are relevant \shortcite{Berges:2010ez}.
The timeline of distributions shown in \Fig{ModeOccupationSequenceEarly} suggests that the evolution of the gas involves transport of particles originating  from the intermediate momentum regime $k\simeq0.05\ldots0.2$, which during the initial evolution gets strongly \emph{overpopulated}.
The particles drift both towards lower and higher wave numbers, building up a bimodal power-law distribution.
To describe the character of this bidirectional flux we plot, in \Fig{PFlux2D3D}, the radial particle and kinetic-energy flux distributions $Q_{k}$ and $P_{k}$, respectively, at $\overline{t}=6550$ corresponding to the bottom center panel of \Fig{ModeOccupationSequenceEarly}. 
Note that the radial particle flux density $Q_{k}$ is multiplied by $gn$ to have the same units as the energy flux density $P(k)$. 
These flux densities are defined through the balance equations \eq{BalEqQ} and \eq{BalEqP}, respectively, with kinetic energy density $\varepsilon_k = n_k k^2/2m$. 
They are determined by integrating the numerically obtained particle and energy spectra $N(k)$ and $E(k)$ up to the scale $k$.

The graph supports the interpretation of the transport in terms of an inverse particle cascade in the IR and a direct energy cascade in the UV  \shortcite{Nowak:2011sk}, in accordance with the appearance of the bimodal momentum distributions in \Fig{ModeOccupationSequenceEarly}. 
Although the derivation of the IR exponents requires the full dynamical theory with non-perturbatively resummed self-energies, the signs of the fluxes correspond to the respective scaling exponents, i.e., $\zeta_{Q}^{\mathrm{IR}}$ in the IR, and $\zeta_{P}^{\mathrm{UV}}$ in the UV.
Moreover, at late times, the kinetic-energy flux $P$ almost vanishes due to a thermalised UV momentum distribution, but $Q$ still reshuffles particles and therefore energy, with the zero mode acting as a sink, keeping the system out of equilibrium close to the non-thermal fixed point. 

 \begin{figure}[!t]
\flushleft
(a) \hspace{0.45\textwidth} (b) \\ \includegraphics[width=0.47\textwidth]{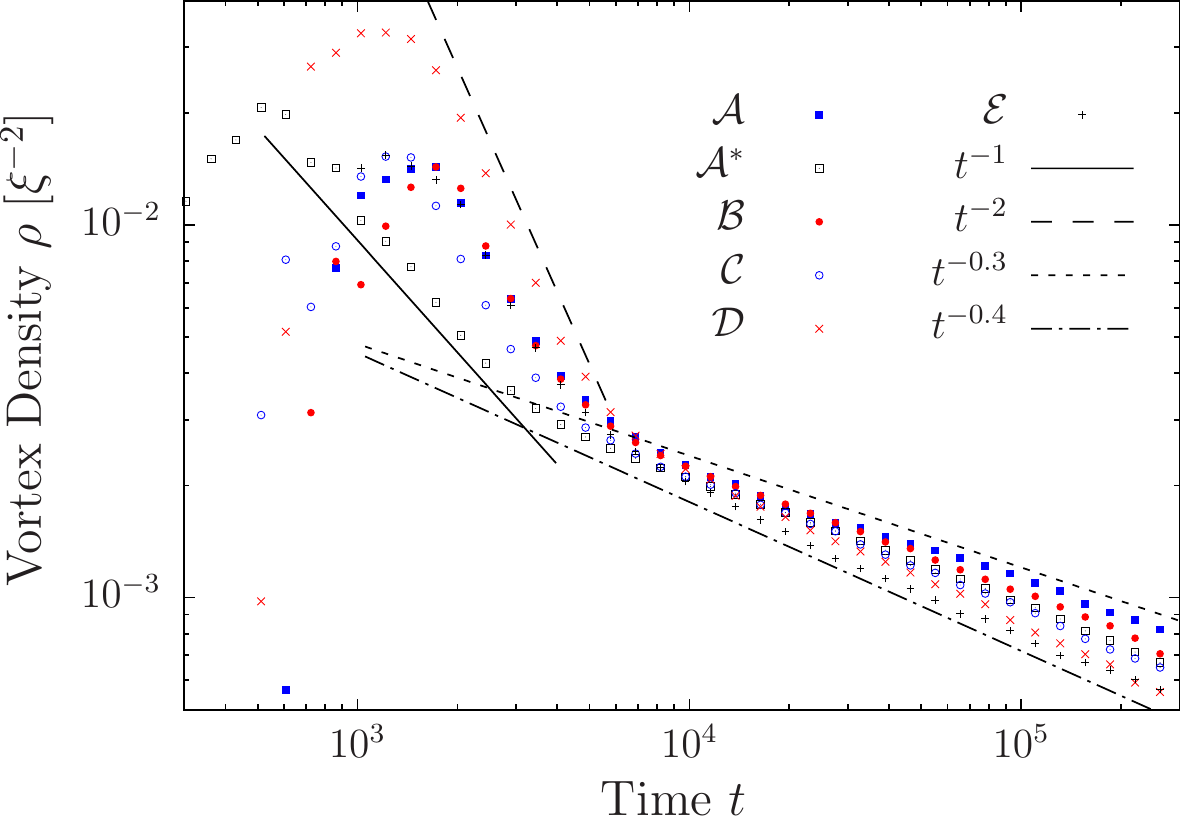}
 \hspace{0.05\textwidth} \includegraphics[width=0.43\textwidth]{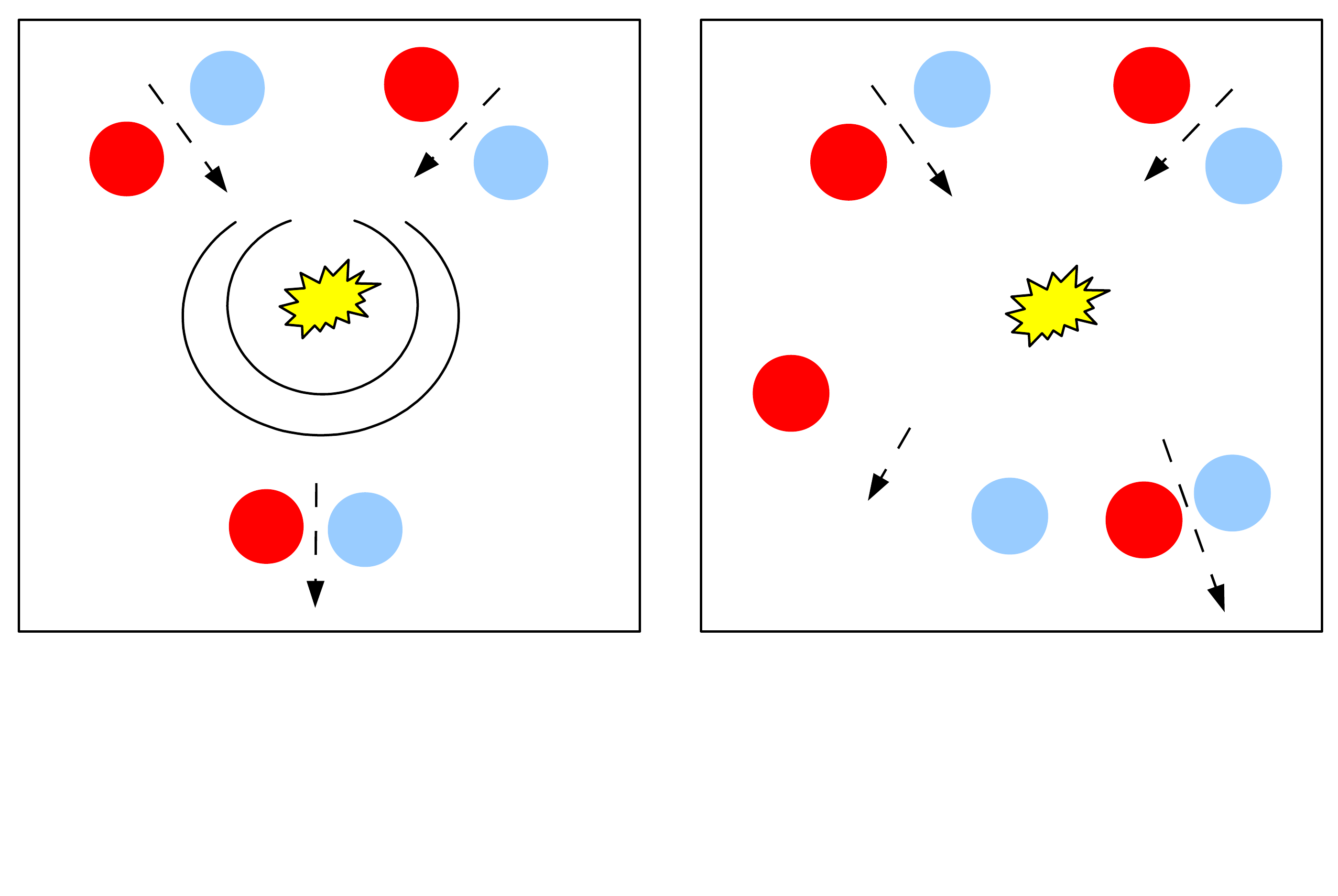} 

 \caption{(a) Vortex density $\rho$ as a function of time $\overline{t}$.
 Evolution for various initial conditions {\protect\shortcite{Schole:2012kt}}, averaged over 20 runs on a grid of size $N_s=1024$. 
 Lines show different power-law evolutions. 
 The vortex density follows power laws $\rho(t)\sim t^{-\alpha_i}$ with two different exponents $\alpha_i$, $i=1,2$. 
 The exponent during the early stage depends considerably on initial conditions $1 \lesssim \alpha_1 \lesssim 2$, whereas the late stage features a
 decay exponent in a narrow interval $0.3 \lesssim \alpha_2 \lesssim 0.4$. 
 The closest approach to the non-thermal fixed point is  at $t\simeq(5\dots10)\times 10^{5}$.
 (b) Sketch of different scattering events between two vortex-antivortex pairs in $d=2$ dimensions. Left panel: Scattering of two vortex pairs resulting in mutual annihilation of two vortices and the emission of density waves. 
 Right panel: Scattering of two vortex pairs, leading to a change in the vortex-antivortex distance $l_\mathrm{D}$ and pair velocity. We remark, that once $l_\mathrm{D} \sim \xi$ a vortex pair decays rapidly under the emission of density waves.}
 \label{fig:VortexCount2D2}
 \label{fig:VortexCountImbalance}
 \label{fig:VortexScattering}
 \end{figure}
We finally remark that a necessary condition for a non-equilibrium stationary distribution is energy damping at large $k$ \shortcite{Zakharov1992a}.
Moreover, energy and particle number conservation in the interaction of different momentum modes can be shown to imply the existence of at least one more sink, i.e., a region where the right-hand sides of \Eq{BalEqQ} effectively has an additional damping term $\sim \Gamma(k)n(k)$, with negative $\Gamma$.
In between these sinks, a source region supplies the input to the bidirectional flux pattern towards the UV and IR.
Using kinetic theory, one can show that under certain conditions a positive $k$-independent flux transports energy, $P>0$, while a negative flux transfers particles, $Q<0$ \shortcite{Zakharov1992a}, 
Remarkably, this pattern remains valid in our case, besides the UV weak-wave-turbulence regime also in the IR region where the exponent emerges from a fixed point of the non-perturbative  dynamic equations for Green's functions.
As already pointed out by~\shortciteN{Scheppach:2009wu}, the derivation of the IR exponent $\zeta_{Q}^{\mathrm{IR}}=d+2$  requires sufficiently well defined quasi particles, suggesting a treatment in terms of the Quantum Boltzmann equation with a momentum dependent scattering matrix element to be applicable.
From this point of view, the negative flux $Q$ and scaling in the IR and the positive flux $P$ and weak wave turbulence in the UV, as observed in the numerics, emerge as a necessary consequence of conservation laws and transport processes described by wave-kinetic transport equations with non-trivial interactions.

\subsection{Approach of the non-thermal fixed point and critical slowing down}

Let us study in some more detail the evolution of the system towards and away from the non-thermal fixed point and focus on universal aspects of the dynamics of vortices. 
For details we refer to \shortcite{Schole:2012kt}.
\Fig{VortexCount2D2}a shows the time evolution of the vortex density
$
 \rho(t) 
 =\langle N^\mathrm{V}(t)+N^\mathrm{A}(t)\rangle/V,
$
where $N^\mathrm{V(A)}(t)$ is the mean number of vortices (antivortices) in the volume $V$ at time $t$, for different specific choices of the initial state, cf. \shortcite{Schole:2012kt}. 
In all runs, vortex formation occurs around ${t}_\mathrm{V}\simeq 10^{3}$, apparent from the steep increase of vortex density around this time. 
For ${t}\gtrsim {t}_\mathrm{V}$, two distinct stages in the vortex density decay are observed, a rapid early stage and a slow late stage. 
We have repeated our simulations on various grid sizes, $N_s \in [256,..., 4096]$. 
Thereby, we found that decay exponents saturate for and above $N_s=512$. We attribute deviations on smaller grids to effects from regular (integrable) dynamics of few-vortex systems~\shortcite{Aref1983a}.
We remark that the onset of the slow decay coincides with the development of a particular scaling behaviour in the single-particle momentum distribution $n(k) \sim k^{-4}$, which by \shortciteANP{Nowak:2010tm} (\citeyearNP{Nowak:2010tm}, \citeyearNP{Nowak:2011sk}) was shown to signal the approach of the non-thermal fixed point and the formation of a set of randomly distributed vortices. 
In this context, the reduction of the vortex density decay exponent, compared to the early stage of rapid decay, is interpreted as due to (critical) slowing down of the nonlinear dynamics near the non-thermal fixed point.

 \begin{figure}[!t]
\flushleft 
(a) \hspace{0.45\textwidth} (b) \includegraphics[width=0.48\textwidth]{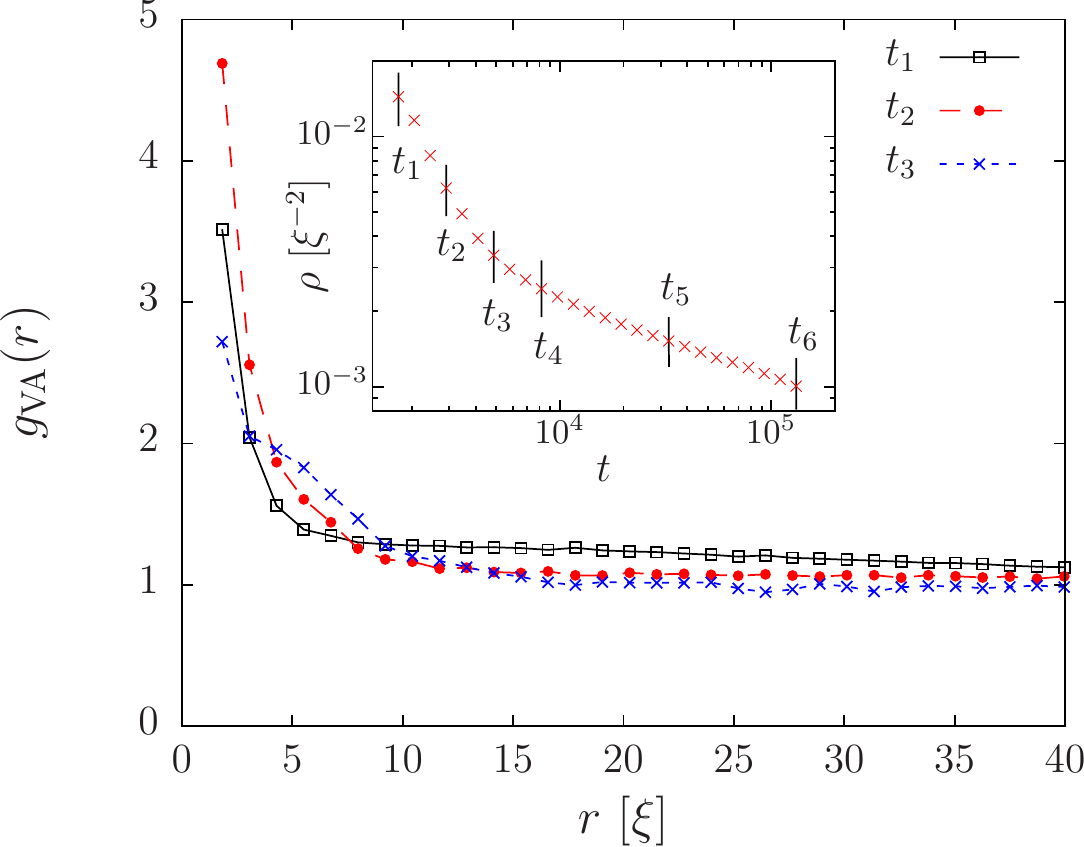}
\hspace{0.01\textwidth} \includegraphics[width=0.48\textwidth]{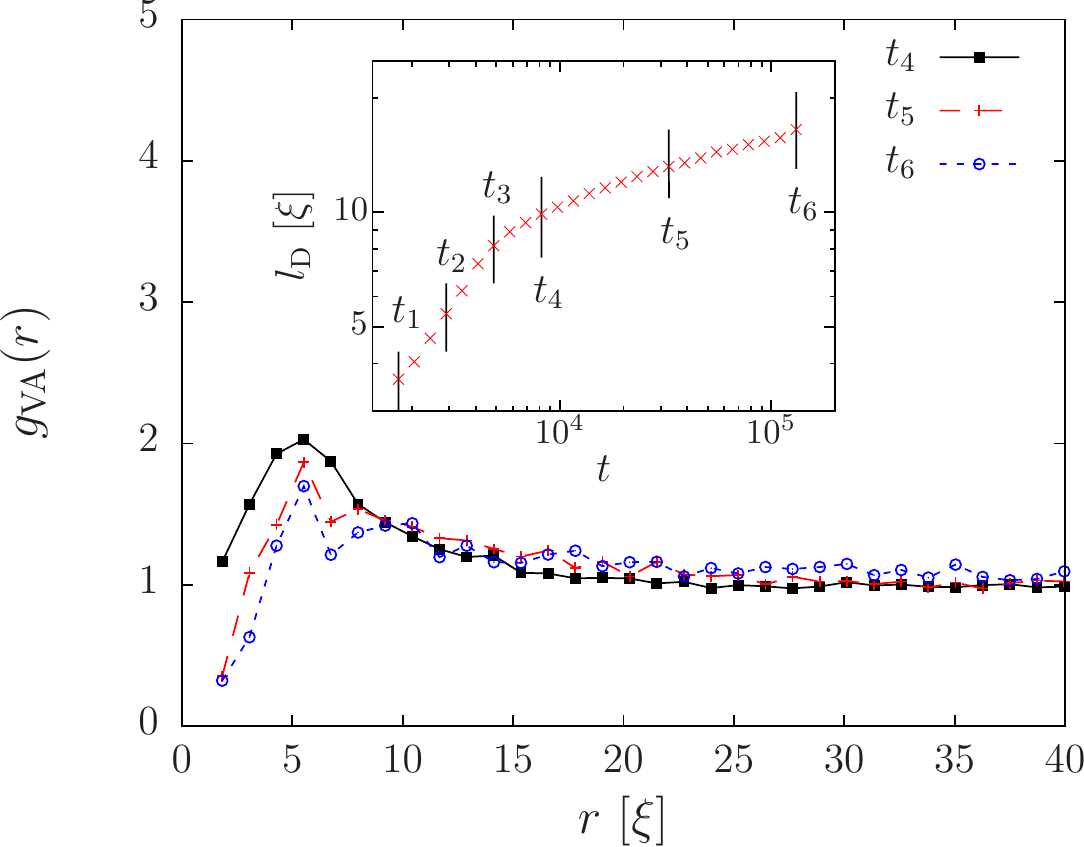}

\caption{Normalised pairing correlation functions $g_\mathrm{VA}$, see \Eq{VortexCorrelation}, as a function of the vortex-antivortex distance $r$, for six different times $t_{i}$ (in lattice units).
(a) $g_\mathrm{VA}(r)$ during the rapid-decay stage, averaged over 174 runs on a grid of $N_s=1024$.
Inset: Vortex density $\rho$  time evolution, from  \Fig{VortexCount2D2}a.
(b) $g_\mathrm{VA}(r)$ during the slow-decay stage, same averaging as in (a). 
Inset: Evolution of mean vortex-antivortex pair distance $l_\mathrm{D}$.
}
 \label{fig:Vortex_Corr_First_Phase}
 \label{fig:Vortex_Corr_Second_Phase}
 \end{figure}

We can discuss the dynamical transition in the vortex annihilation dynamics in terms of characteristic features of the vortex-antivortex correlation function
\begin{equation}
 g_\mathrm{VA}(\vector{x},\vector{x}',t) 
 = \frac{\langle \rho^\mathrm{V}(\vector{x},t)\rho^\mathrm{A}(\vector{x}',t) \rangle}
            {\langle \rho^\mathrm{V}(\vector{x},t) \rangle \langle \rho^\mathrm{A}(\vector{x}',t) \rangle } \, , 
\label{eq:VortexCorrelation}
\end{equation}
where $\rho^\mathrm{V(A)}(\vector{x},t)=\sum_{i}\delta(\mathbf{x}-\mathbf{x}_{i}(t))$ is the distribution of (anti)vortices in a single run at time $t$.
For sufficiently large ensembles, $ g_\mathrm{VA}$ is a function of $r=|\vector{x}-\vector{x'}|$ only.
 At early times, one finds a strong pairing peak in $g_\mathrm{VA}(r,t)$ near $r=0$, see \Fig{Vortex_Corr_First_Phase}a. 
This peak gets quickly reduced and a hole is \textquoteleft burned\textquoteright~into the correlation function near the origin, see~\Fig{Vortex_Corr_Second_Phase}b. 
Following the time evolution of the spatial vortex distribution we observe that this involves qualitatively different processes:
Mutual annihilations of closely positioned vortices and antivortices occur under the emission of sound waves.
Further separated vortices can approach each other in different ways as illustrated in \Fig{VortexScattering}b. 
The scattering of two pairs can directly lead to the annihilation of one pair under the emission of sound waves. 
This includes events where the dipole length reduces below a certain threshold, implying a density dip rather than a vortex-pair. 
This density dip can still interact with other vortices but  will quickly vanish.
Alternatively, the scattering reduces the vortex-antivortex separation within one pair while it increases it within the other, in accordance with the Onsager point-vortex model \shortcite{Onsager1949a}.  
We refer to this characteristic change in $g_\mathrm{VA}(r)$ as a vortex unbinding process. 
Recall that the Onsager model does not contain a kinetic term for the point defects such that their interaction potential energy can not induce relative acceleration of the vortices. 
Changes in their separation must occur dynamically via collisions.

At around the time ${t}_3 \lesssim {t} \lesssim {t}_4$ the power-law exponent of the vortex density decay changes to about a third of its previous value, see the inset of \Fig{Vortex_Corr_First_Phase}a. 
Computing the mean vortex-antivortex pair distance $l_\mathrm{D}$, by averaging over distances between each vortex and its nearest antivortex, we find that, in accordance to the previous discussion, $l_\mathrm{D}$ grows continuously, exhibiting two characteristic stages, see the inset of \Fig{Vortex_Corr_Second_Phase}b. 
At times ${t} \gtrsim 10^4$, $l_\mathrm{D}(t)$ approaches the power-law solution $l_\mathrm{D} \sim \rho^{-1/2}$, as expected for uncorrelated vortices. 
%

    \begin{figure}[!t]
 \center
 \includegraphics[width=0.6\textwidth]{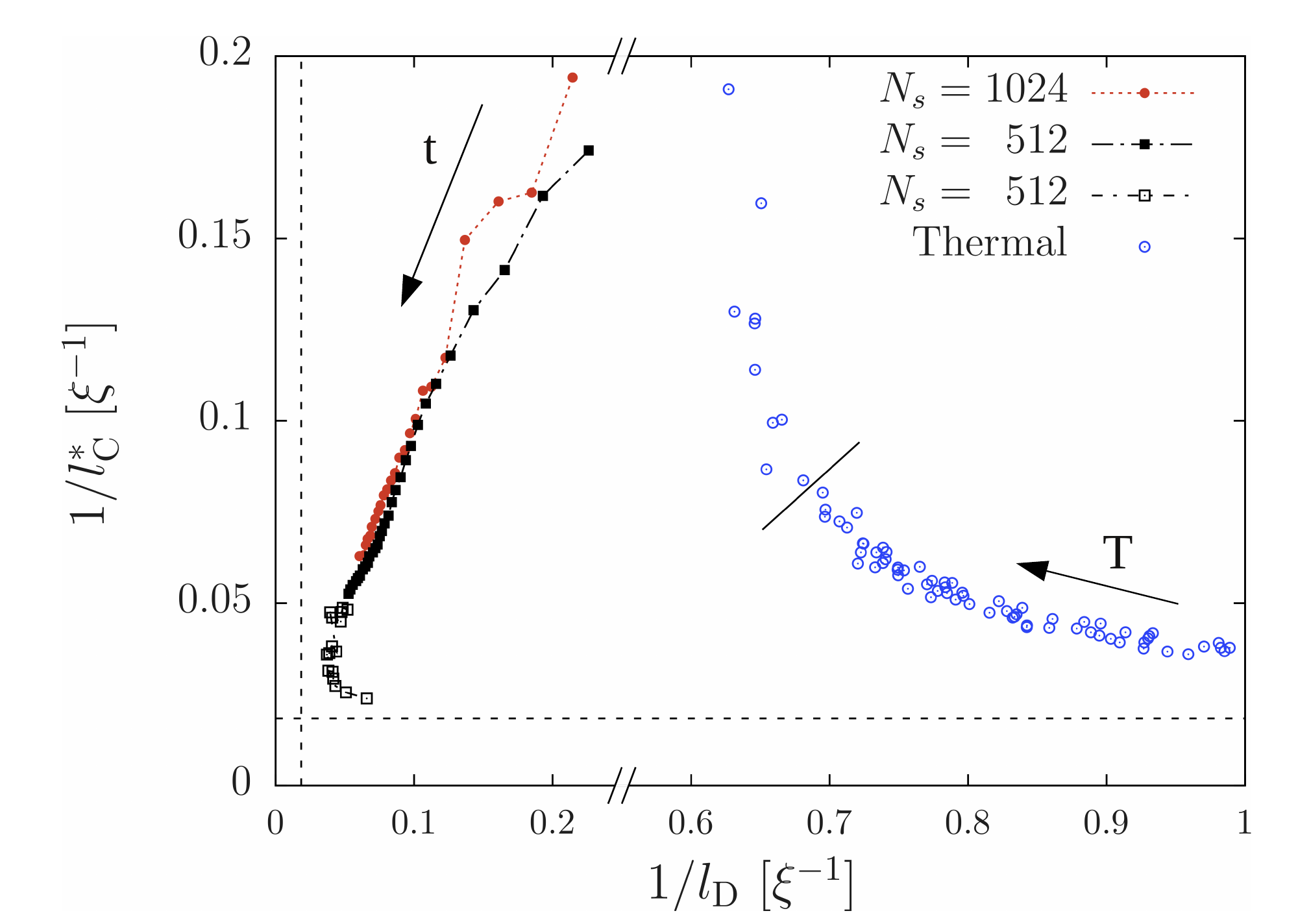}
\caption{Trajectories of multi-vortex states in the space of inverse coherence length $1/l_\mathrm{C}$ and inverse mean vortex-antivortex distance $1/l_\mathrm{D}$, starting from $t=t_\mathrm{V}$. Comparison of the thermal line $(l_\mathrm{D}^{-1}(T), l_\mathrm{C}^{*-1}(T))$ for a range of temperatures $T$ (right) with the corresponding dynamical evolution (left). 
Dashed lines mark the minimal values $2/L=0.018\xi^{-1}$, available on a grid of size $N_s=512$. Note that the $(1/l_{\mathrm{D}})$-axis interval [0.25,0.55] has been cut out.}
    \label{fig:non-thermal fixed pointvsThermal}
    \label{fig:Correlations2D}
    \end{figure}
%

We close by focusing on the growth of long-range coherence, associated with the annihilation of topological defects \shortcite{Levich1978a,Kagan1992a,Kagan1994a,Damle1996a,Berloff2002a,Svistunov2001a,Kozik2009a,Nazarenko2006a}. 
From this point of view, freely decaying superfluid turbulence is a particular example of phase-ordering dynamics after a quench into the ordered phase \shortcite{Bray1994a}. 
Whereas in three dimensions, a second-order phase transition connects a normal-fluid and a superfluid phase, a Bose gas in two dimensions experiences a Berezinskii-Kosterlitz-Thouless (BKT) transition \shortcite{Berezinskii1971a,Kosterlitz1973a}. 
For the two-dimensional ultracold Bose gas, experimental and theoretical results support the understanding of the phase transition in terms of vortices undergoing an unbinding-binding transition \shortcite{Hadzibabic2006a,Simula2006a,Schweikhard2007a,Giorgetti2007a,Weiler2008a,Bisset2009a,Foster2010a}. 

In this context, we are interested in a comparison between correlation properties observed in the non-equilibrium dynamics near a non-thermal fixed point and those known from equilibrium studies. 
We compute the dynamical trajectory of the vortex gas in the space of inverse coherence length and inverse mean vortex-antivortex pair distance. 
We compare our results to simulations of a thermal two-dimensional Bose gas specifically for our system parameters. 
We define a coherence length $l^{*}_\mathrm{C}$ in terms of the integral over the angle-averaged first-order coherence function,
\begin{align}
l^*_\mathrm{C}= \int \mathrm{d}r \,g^{(1)}(r),
\qquad
g^{(1)}(r) &= \int \mathrm{d}\theta \, \frac{ \langle \phi^*(\vector{x})\phi(\vector{x}+\vector{r}) \rangle} {\sqrt{\langle n(\mathbf{x})\rangle\langle n(\mathbf{x}+\mathbf{r})\rangle} } \,.
\label{eq:Coherence}
\end{align}
$l^{*}_\mathrm{C}$ measures the spatial extension of the first-order coherence function. 
Contrary to $r_\mathrm{coh} = \int \mathrm{d}{r} \, r^{2} g^{(1)}(r)/\int \mathrm{d}{r}\, r\,g^{(1)}(r)$, the quantity $l^*_\mathrm{C}$ does not sum up values of $g^{(1)}(r)$ weighted by the distance, which would enlarge insignificant contributions at large $r$. 
In addition, it does not overestimate the coherence of flat distributions.
In equilibrium this quantity smoothly interpolates between the regime of exponential decay of $g^{(1)} \sim \mathrm{exp}( -r/\xi_\mathrm{C} )$ above the BKT transition
and its power-law decay below. 

In \Fig{Correlations2D}, we follow the time evolution of the gas for ${t} >{t}_\mathrm{V}$. 
One can observe that a state of low coherence and small mean vortex-antivortex pair distance evolves towards larger coherence and larger vortex-antivortex separation. 
As discussed above, this is due to vortex annihilations and vortex-antivortex unbinding. For times ${t} > 10^4$, the coherence length grows as $l^*_\mathrm{C} \sim \rho^{-1/2}$, in the same way as $l_\mathrm{D}$ shown in \Fig{Vortex_Corr_Second_Phase}b. 
The evolution considerably slows down for $1/l^*_\mathrm{C} \sim 1/l_\mathrm{D} \rightarrow 0$, when the gas starts to show the characteristic scaling $n(k)\sim k^{-4}$.
After spending a long time near the non-thermal fixed point, $l_\mathrm{D}$ declines because the last remaining vortex-antivortex pairs reduce their size prior to their annihilation and the equilibration of the system.
Our understanding of the non-thermal fixed point as a configuration with a few, maximally separated pairs on an otherwise maximally coherent background implies it to be located near the crossing of the dashed lines. 
Hence, the non-thermal fixed point is approached most closely between $t\simeq 5\times 10^{5}$ and $t\simeq10^{6}=t_{3}$.
To set the above evolution in relation to equilibrium configurations, we also show the thermal line $\{l_\mathrm{D}^{-1}(T), l_\mathrm{C}^{*-1}(T)\}$ for a range of temperatures $T$ for which the zero-mode population does not vanish. 

Our results allow to draw a picture of the evolution path towards and away from the fixed point. 
The non-thermal fixed point is characterised by a few pairs -- in the extreme case one pair --  of far-separated anti-circulating vortices and bears similarities with the equilibrium BKT fixed point.
However, while the phase transition also features unbinding of vortices, finite temperature implies the simultaneous excitation of many rotons, i.e. strongly bound vortex-antivortex pairs.
The non-thermal fixed point is clearly identified by strong wave turbulent scaling in the infrared limit, $n(k)\sim k^{-4}$. 
The high-energy modes are much weaker populated, e.g., far below the BKT critical temperature or remain out of equilibrium.
The details of the UV mode populations are determined by the way the non-thermal fixed point is being approached.

The way the system is forced, here, to approach the non-thermal fixed point generalises the protocol of \shortciteN{Kibble1976a}  and \shortciteN{Zurek1985a}.  
A strong sudden quench replaces the adiabatic approach of the BKT transition. 
The dynamical evolution in the vicinity of the BKT critical point was studied by \shortciteN{Mathey2009a}, in terms of a perturbative renormalisation group analysis. 
The route to a non-perturbative analysis in the strong-coupling regime is provided by out-of-equilibrium functional techniques, see Refs.~given at the end of \Sect{IRScalingAsSWT}.

\subsection{Dependence on driving and a new route to Bose condensation}

\begin{figure}[!t]
\center
\includegraphics[width=0.75\textwidth]{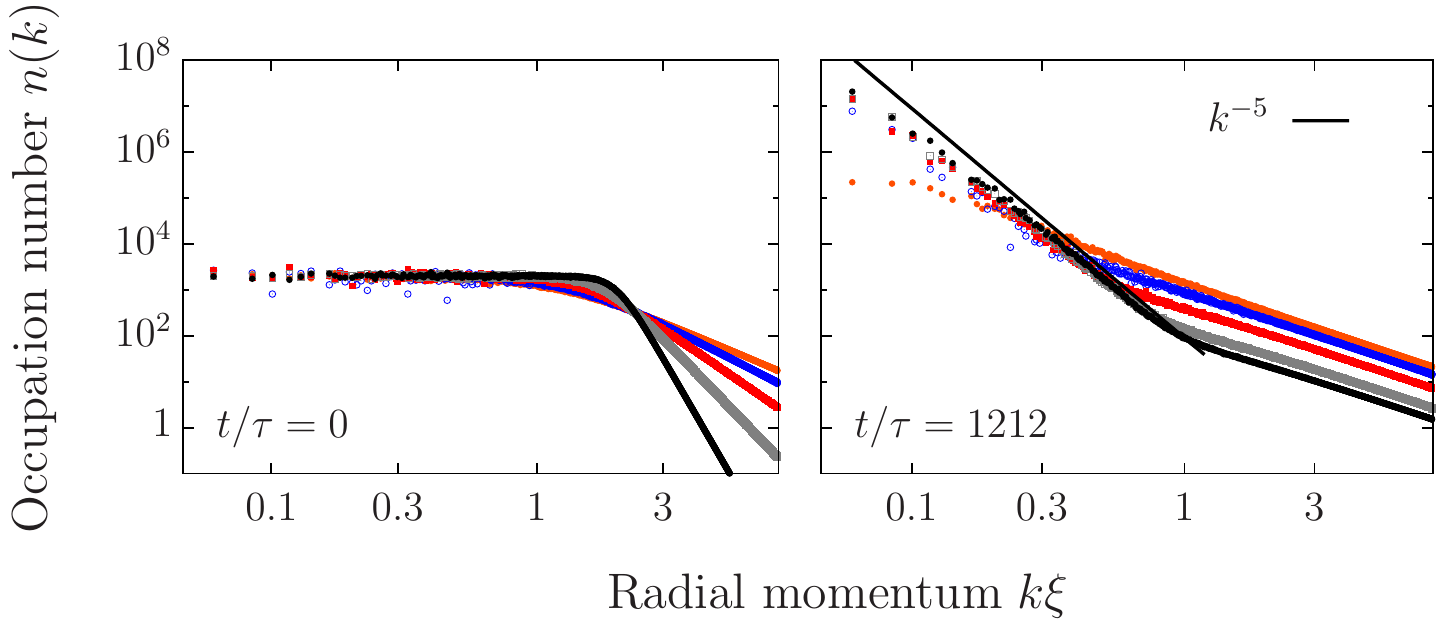}\vspace{-0.02\textwidth}
\caption{Momentum distributions of a 3-dimensional gas for different initial conditions. Left: Initial conditions with varying high momentum decay $k^{-\alpha}$, characterised by the paramter $\alpha=2.5,3,4,6,10$, representing different evaporative cooling quenches. Right: At intermediate times, when the system is closest to the non-thermal fixed point, showing the steep IR power laws $\sim k^{-5}$ for initial conditions representing a strong cooling quench, $\alpha \gtrsim 3$.}
\label{fig:Evaporative}
\end{figure}
%
Let us finally give a taste of the relevance of the strength with which the system is driven away from thermal equilibrium for the approach of the non-thermal fixed point and, as a by-product, find new aspects of how superfluid turbulence affects the process of Bose condensation  \shortcite{Nowak:2012gd}.
In this section we restrict ourselves to a gas in $d=3$ dimensions.
We choose the overpopulated momentum distribution of the gas directly, eliminating the early instability phase of our previous simulations during which overpopulation is induced through non-linear scattering of wave modes.
The initial field in momentum space, $ \phi(\mathbf{k},0) = \sqrt{n(\mathbf{k},0)} \mathrm{exp}\{ i \varphi (\mathbf{k},0) \} $,  is parametrized in terms of a randomly chosen phase $\varphi(\mathbf{k},0) \, \in \, [0,2\pi)$ and a density $n(\mathbf{k},0) = f(k)\nu_\mathbf{k}$, with $\nu_\mathbf{k}\ge0$ drawn from an exponential distribution $P(\nu_\mathbf{k})=\exp(-\nu_\mathbf{k})$ for each $\mathbf{k}$.
The spectrum is flat at low $k$ and falls off according to the function
$
 f(k) = {f_{\alpha}}/({k_{0}^{\alpha}+k^{\alpha}}),
$
where $\alpha$ controls the deviation from a thermal decay with $\alpha=2$.
We choose a cutoff $(k_{0}\xi)^{\alpha}=0.2/0.44^{\alpha}$ and normalization $f_{\alpha}=400/0.44^{\alpha}$.
We compare results for a range of different cooling quenches defined by the power-laws $\alpha=2.5,\dots,10.0$, varying the total number between $N=10^{9}$ ($\alpha=2.5$) and $N=4.3\times10^{8}$ ($\alpha=10$).

In \Fig{Evaporative} we show $n(k,t)$ over the radial momenta $k=|\mathbf{k}|$ at the initial time as well as at a moment when the system is closest to the non-thermal fixed point, for the different choices of $\alpha$. 
During the initial evolution ($t \lesssim 10^2\tau$) the mode occupations gradually spread to lower wave numbers, at the same time depositing energy into the high-momentum tail.
We emphasise that cutting away sufficiently much population at high momenta initially is necessary if the system is supposed to approach the non-thermal fixed point during its rethermalisation:
As described in \Sect{Fluxes}, the approach of the fixed point is characterized by a dual cascade  in which the energy of the intermediate-$k$ overpopulation gets deposited in the high-$k$ tail, carried there by a few particles,  while the majority of overpopulation particles moves towards the IR, conserving overall energy and particle number. 
Hence, only strong cooling quenches allow for the build-up of a steep population far into the IR.
 
\begin{figure}[!t]
\center
\includegraphics[width=0.9\textwidth]{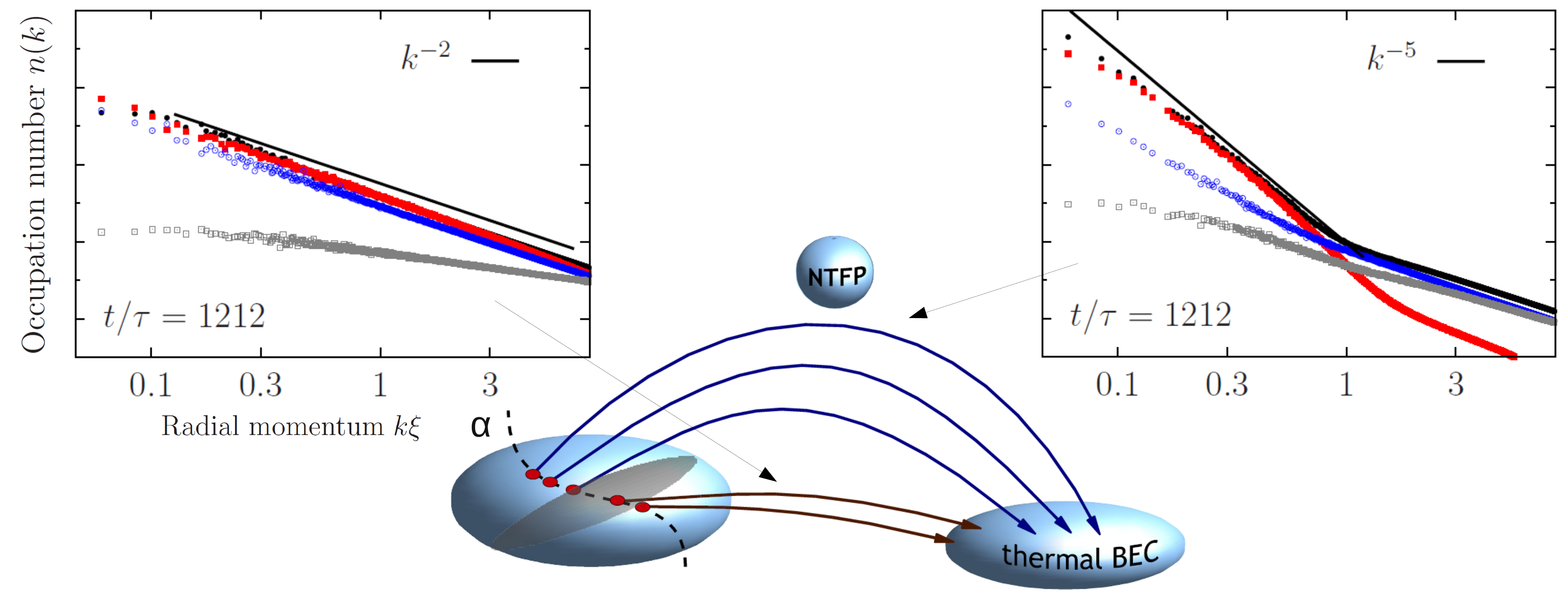}
\caption{Depending on the strength $\alpha$ of the initial cooling quench, the gas can thermalise in a near-adiabatic manner to a Bose-Einstein condensate.
Alternatively, it can first approach and be critically slowed down near a non-thermal fixed point where it is characterised by a scaling spectrum $n(k)\sim k^{-5}$ in the IR.
Furthermore, dynamical scale separation of the incompressible (red points) and incompressible (blue points) components of the gas (cf.~\protect\shortciteN{Nowak:2011sk}, and last par.~of \protect\Sect{VorticesinaSF}) occurs.
The incompressible component corresponds to transverse vortical flow.
 The grey points indicate the quantum-pressure component.
}
\label{fig:NTFPScheme}
\end{figure}
%
At late times, the spectra developing from the different initial $\alpha$ differ strongly. 
For $\alpha\gtrsim3$, the distribution develops a bimodal structure, with a power-law behaviour $n(k) \sim k^{-5}$ in the infrared (IR) and $n(k) \sim k^{-2}$ in the UV. 
At very long times, this bimodal structure decays towards a global $n(k) \sim k^{-2}$ (not shown). 
For $\alpha \lesssim 3$, the distribution directly reaches a thermal Rayleigh-Jeans scaling $n(k)\sim T/k^{2}$. 
Preliminary results \shortcite{Nowak2013a} show that the trajectories, when plotted as in \Fig{Correlations2D}, approach the non-thermal fixed point in the lower left corner the closer, the larger $\alpha$ is chosen.
While $\alpha=2.33$ leads to a trajectory near the thermal states, $\alpha=10$ induces a motion similar to the one of the black points.

Much experimental effort is undertaken at present to study the dynamics of condensation under conditions of rapid evaporative cooling \shortcite{Pitaevskii2003a,Ritter2007a,Weiler2008a,Smith2011a}.
Moreover, many-body dynamics of coherent bosonic excitations is intensively studied in solid-state systems consisting of magnons~\shortcite{Demidov2007a,Demidov2008a,Keeling2008a,Nowik2012a} or polaritons~\shortcite{Saba2001a,Kasprzak2006a,Balili2007a,Lagoudakis2008a,Amo2011a,Keeling2011a}. 
Recently, condensation has been discussed for the case of gluons as an intermediate stage of heavy-ion collisions~\shortcite{Blaizot:2011xf,Berges:2012ks} and, for relativistic scalars, been found to rely on a nonperturbative inverse particle cascade \shortcite{Berges:2012us}.

Our results confirm that Bose-Einstein condensation in a non-equilibrated and under-cooled gas can have the characteristics of a turbulent inverse cascade \shortcite{Svistunov1991a,Kagan1992a,Kagan1994a,Berloff2002a}, corresponding to a quasi-local transport process in \emph{momentum} space, into the low-energy modes of the Bose field. 
The main new finding is that the superfluid turbulence period can appear in two  different forms~\shortcite{Nowak:2012gd}. 
The two possible paths to Bose-Einstein condensation are shown schematically in \Fig{NTFPScheme}. 
If a sufficiently small amount of energy is removed, a thermal Rayleigh-Jeans distribution forms in a quasi-adiabatic way. 
The chemical potential increases, and a fraction of particles is deposited in the lowest mode, forming a Bose-Einstein condensate. 
In the second scenario, after a sufficiently strong cooling quench, the system develops transient scaling behaviour in the momentum distribution prior to condensate formation, i.e., it approaches the non-thermal fixed point. 
These scenarii differ qualitatively in how the condensate mode builds up as a function of time~\shortcite{Nowak:2012gd}.

\section{Other systems}

\subsection{Soliton ensembles}

\begin{figure}[!t]
\center
\includegraphics[width=0.5\textwidth]{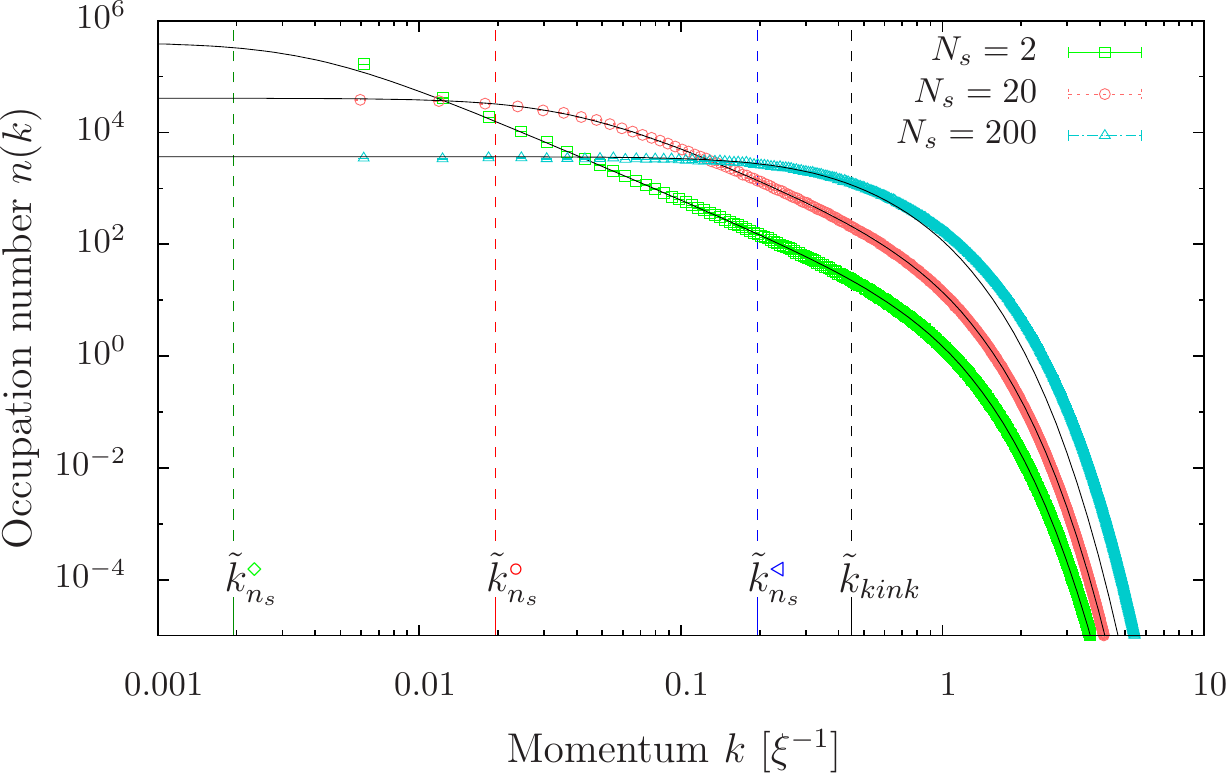}
\caption{Momentum distributions for $N_s$ randomly distributed black, i.e., static solitons (symbols), showing a $k^{-2}$ power law behaviour between the scale $k_{n_s}$ given by the mean soliton separation and $k_{kink}$ given by the healing length. The spectra are found to be in excellent agreement with analytically predicted lines \protect\shortcite{Schmidt:2012kw}.
}
\label{fig:SolitonSpectrum}
\end{figure}
%
In the preceding sections, it was discussed how strong wave turbulent scaling of the momentum spectrum of a degenerate Bose gas can be understood from the statistics of vortices. 
In this context, correlations between vortices and antivortices play a crucial role.
In three spatial dimensions similar relations exist between scaling and the creation of vortex lines and rings \shortcite{Nowak:2011sk}. 
Looking at systems in one spatial dimension we find that solitons, in particular dark solitons play a crucial role in realising strong IR wave turbulence. 
Solitons interact with other defects as well as sound excitations and show characteristics of localised quasi-particle excitations. 
In the following we briefly summarise to what extent random ensembles of solitons can be seen as characterising a non-thermal fixed point in one-dimensional bosonic systems.

\begin{figure}[!t]
\center
\includegraphics[width=0.8\textwidth]{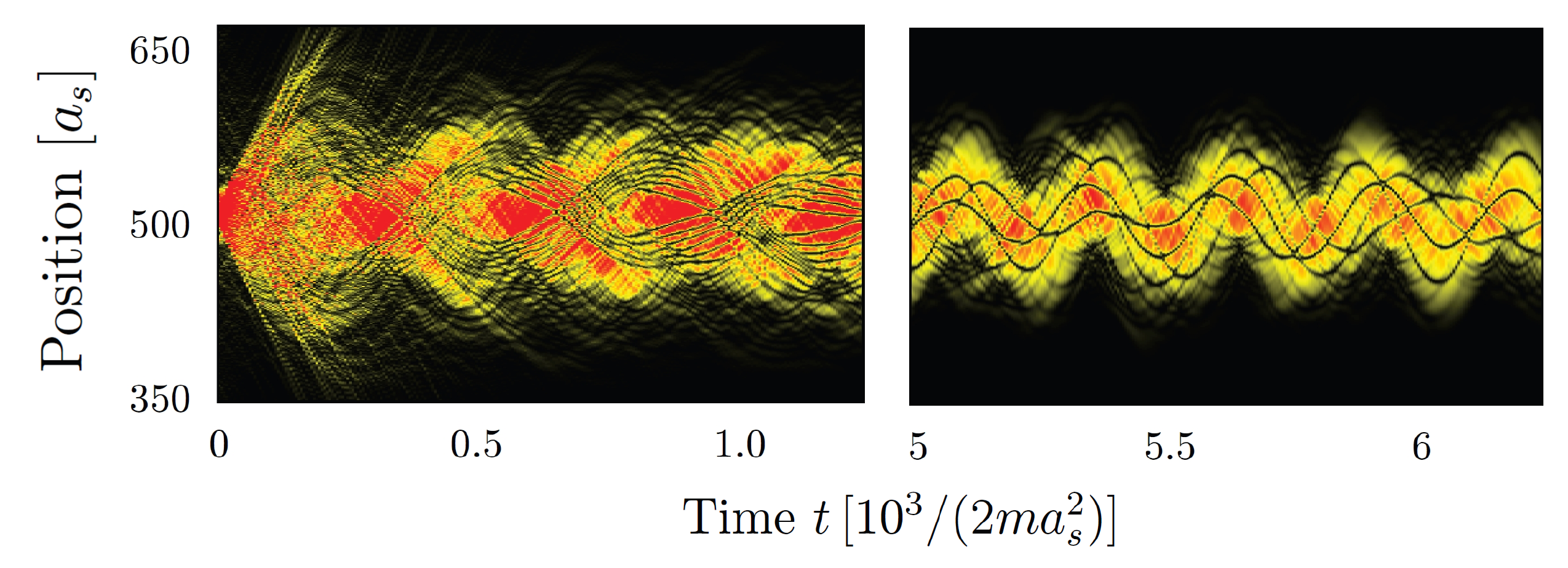}
\caption{Time intervals of a single run of the classical field equation \eq{GPE}, showing solitons which oscillate inside a trapped one-dimensional ultracold Bose gas. 
The gas is initially noninteracting and thermalised, with $T=360\omega_{\mathrm{ho}}$, in a trap with oscillator length $l_{\mathrm{ho}}=8.5$ (in grid units). 
At time $t=0$ the interaction is switched to $g_{\mathrm{1D}}=7.3\times10^{-3}$, and a cooling period using a high-energy knife is applied.
The panels show the one-dimensional colour-encoded density distribution as a function of time. 
Left panel: The initially imposed interaction quench causes strong breathing-like oscillations and the creation of many solitons. 
Right panel: Breathing oscillations have damped out, leaving a dipolar oscillation of the bulk distribution, with clearly distinct solitons in the trap. 
For details cf.~\protect\shortciteN{Schmidt:2012kw}.
}
\label{fig:SolitonDynamics}
\end{figure}
%

A model of randomly positioned grey solitons, being solutions of the Gross-Pitaev\-skii equation \eq{GPE} in the vicinity of each soliton density dip, was discussed by \shortciteN{Schmidt:2012kw}. 
In the limit of large soliton separations compared to the healing length, analytic expressions for the momentum spectra were derived in a homogeneous system as well as under the constraint of a trapping potential. 
A central result is depicted in \Fig{SolitonSpectrum} for the case of randomly distributed black solitons in a homogeneous background. 
The momentum distributions show a $k^{-2}$ power law behaviour between the scale $k_{n_s}$ corresponding to the mean soliton separation and $k_\mathrm{kink}$ which marks the healing length scale. 
At higher momenta the distribution features an exponential decay following from the characteristic way the spatial density drops inside the soliton cores while at low momenta a flat distribution marks the long-range exponential decay of the coherence.
The soliton configurations in one dimension resemble the turbulent phenomena in $d=2$ and $3$ dimensions. 
Their presence is accompanied by a quasi-stationary power-law momentum distribution which marks the self-similarity of a random distribution of sharp phase kinks:
looking at the system within a window, the size of which is below the mean soliton separation and well above the core width, the system looks the same irrespective of the window size. 
There is either a kink seen within the window or not.

Transport in momentum space, in the form of an inverse cascade towards the IR, corresponds to soliton decay which increases the inter-soliton distance and extends the self-similar regime towards smaller momenta (see \Fig{SolitonSpectrum}). 
We remark that the predicted strong wave turbulent scaling $\sim k^{-d-2}$ which was found consistent with the vortex scaling in $d=2$ and $3$ does not give the scaling $\sim k^{-2}$ for the solitons in $d=1$.
The reason for this is expected to be similar as for the case of domain walls in $d=2, 3$, in a multi-component system as described in \Sect{SpinGases} below:
new transport equations must be set up in which a different conserved current, relating to transport of spin wave excitations, leads to a different IR scaling. 
A soliton in $d=1$ does not exhibit transverse (incompressible) superfluid flow as that around a vortex core  which decays as $|\mathbf{v}(r)| \sim 1/r$. 
In this context we remark that the longitudinal (compressible) component in $d=2$ and $3$ dimensions (\Fig{NTFPScheme}) shows a by one weaker IR exponent. 

The formation and evolution of soliton excitations in trapped one-di\-men\-sio\-nal Bose gases can be studied by means of the classical field equation. 
A possible scenario of far-from-equilibrium dynamics involves  an initially non-interacting thermal gas that is quenched by a sudden ramp of the interaction, as studied by~\shortciteN{Schmidt:2012kw}. 
To allow the emerging collective excitations to form solitons at a desired density, evaporative cooling helps to achieve the required reduction of the UV mode populations. 
An exemplary position-space evolution is depicted in \Fig{SolitonDynamics}.

\subsection{Domains and defects in two-component systems}
\label{sec:SpinGases}

\begin{figure}[!t]
\includegraphics[width=1.00\textwidth]{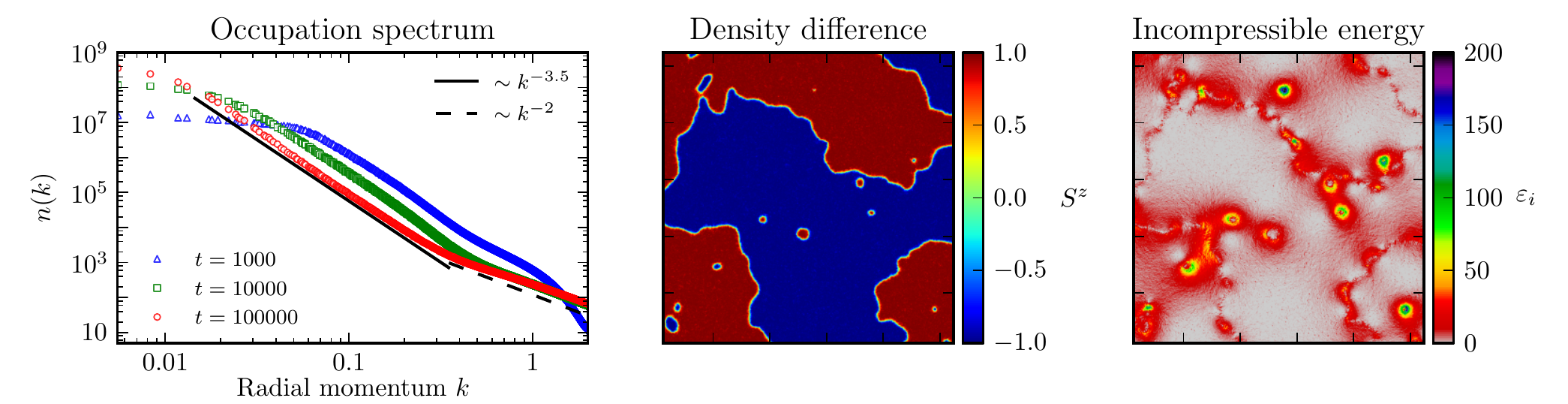}
\caption{Non-thermal fixed point and domain formation in an immiscible two-component Bose gas. 
Left: Occupation number spectrum in $d = 2$ spatial dimensions at different times (note double-log scale). 
The spectrum at late times shows a bimodal structure with a thermal scaling $n(k) \sim k^{-2}$ at high momenta and an IR power-law 
$n(k) \sim k^{-3.5}$, similar to scaling seen when charge separation is found in the relativistic simulations, see \protect\Fig{charge}. 
Centre: Polarization $\langle S_{z}(\vector{x})\rangle$, i.e., density difference of the two gas components, $S^z(\vector{x}) = (n_1 - n_2)/(n_1 + n_2)$. It demonstrates the formation of spin domains as a result of dynamic demixing. Right: Spatial distribution 
of incompressible kinetic energy, $\varepsilon_i(\vector{x})$, showing high amounts of transverse hydrodynamic, i.e., vortical flow around spin domains and especially around point-like defects.}
\label{fig:spinorgas}
\end{figure}

As a further example we briefly comment on what is already known in the context of strong wave turbulence and defect formation in multi-component, e.g., spinor gases~(\shortciteN{Karl2012a}; \Fig{spinorgas}). 
A two-component Bose gas with contact interactions is well known to possess two different ground states depending on the value of the parameter $\alpha=g_{11}g_{22}/g_{12}$ which parametrizes the relative strength of intra-species couplings $g_{11}$ and $g_{22}$ as compared to inter-species interaction $g_{12}$~\shortcite{Timmermans1998a,Kasamatsu2006a} and is the subject of increasing experimental investigations \shortcite{Sadler2006a,Vengalattore2008a,Guzman2011a,Nicklas2011a}. 
For $\alpha > 1$, which is called the immiscible regime, the inter-species interaction energy is greater than that of the interactions within one species. Hence, in the ground state of the system, the spatial overlap of the components is minimized through domain formation and spatial separation of the two components. 
In contrary, for $\alpha<1$, the two components become miscible and uniformly distributed when in the ground state. 
One can  use the parameter $\alpha$ to change the properties of the system in the yet unexplored region of non-equilibrium quasi-stationary states. 

%
Dynamical instabilities serve to drive the system far from equilibrium. 
This leads to momentum distributions of the different components which are characterized by a strong overpopulation at intermediate momenta as compared to thermal equilibrium distributions with the same energy and particle numbers. 
Similar to our finding for the case of a one-component Bose gas in $d=1,2,3$ dimensions, such a far-from-equilibrium distribution of particle momenta induces a redistribution of particles in momentum space both towards lower and higher momenta. 
Subsequently, the system encounters long-lived transient states with non-topological  and quasi-topological defects including domain walls, vortices in a single species, and skyrmions in the coupled spin system. 
Distinguishable types of defects are produced for different values of the external parameter such that this allows to induce a transition between meta-stable non-equilibrium ordered states. 
In summary, one obtains an example of how to extend the concept of a phase transition into the realm of far-from-equilibrium time-evolution.

\subsection{Charge separation in reheating after cosmological inflation}

In cosmological models of the universe, reheating describes the epoch starting at the end of inflation \shortcite{Allahverdi:2010xz}.  
During this period the potential energy of the inflaton field is redistributed into a homogeneous and isotropic hot plasma of particle excitations. 
These become a substantial part of the further expanding universe.  
Simple models describing reheating after inflation invoke self-interacting scalar fields.  
One of the popular scenarios involves the parametrically resonant amplification of quantum fluctuations of the macroscopically oscillating inflaton field.  
The amplified modes represent the emerging matter content of the universe \shortcite{Kofman:1994rk,Traschen:1990sw}. 
Various theoretical approaches have been proposed to model reheating. 
As both, the inflaton and the amplified modes are strongly populated, classical field simulations can be applied to describe their evolution \shortcite{Khlebnikov:1996mc,Prokopec:1996rr,Tkachev:1998dc}.

\begin{figure}[!t]
\includegraphics[width=0.368\textwidth]{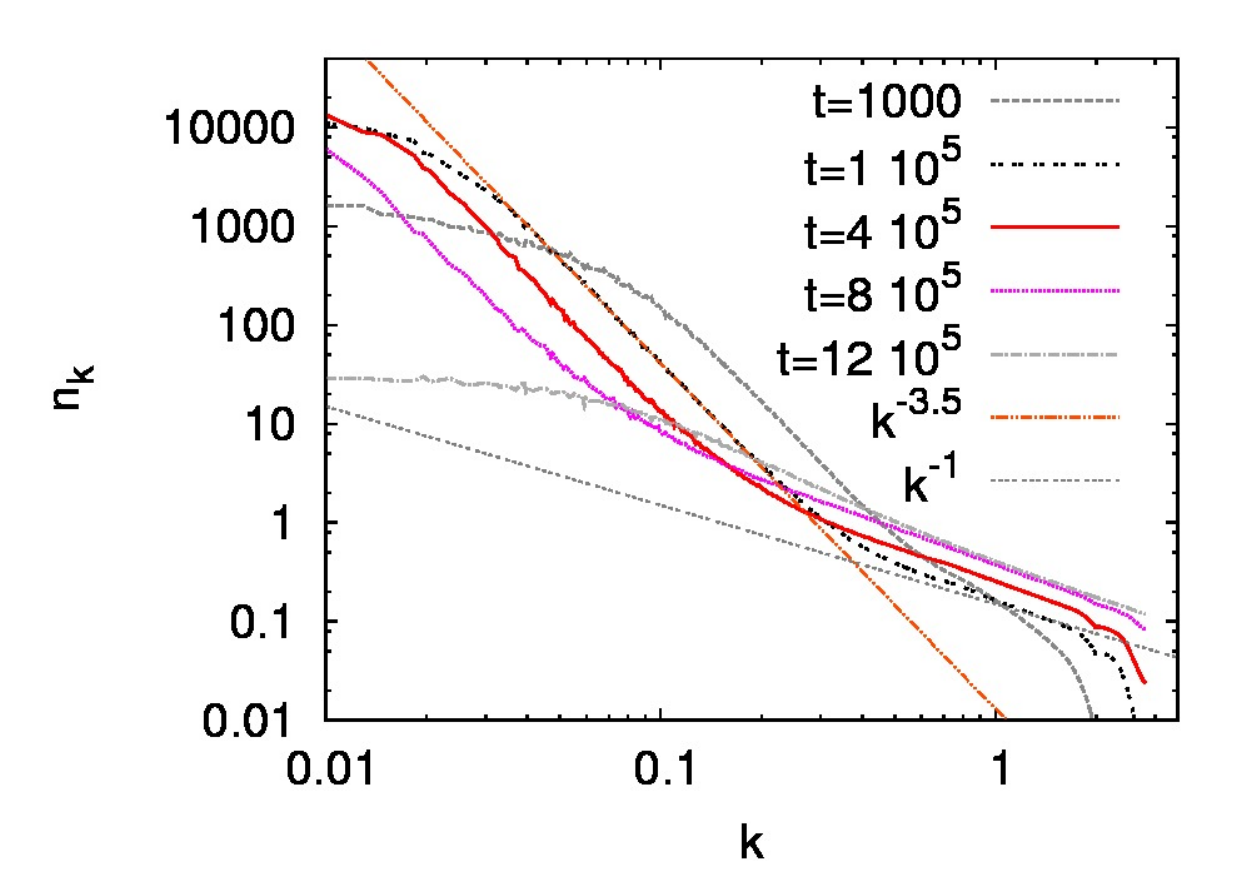}
\ 
\includegraphics[width=0.292\textwidth]{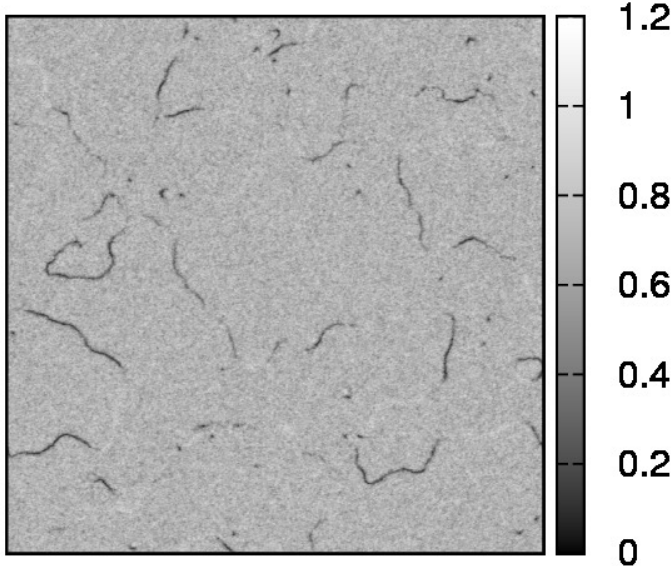}
\
\includegraphics[width=0.292\textwidth]{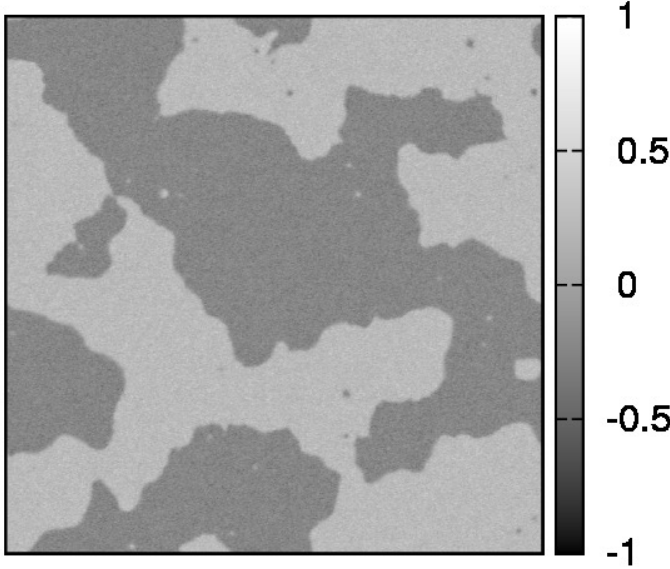}
\ \\
\caption{Non-thermal fixed point and charge separation in the nonlinear Klein-Gordon equation for a complex field. 
Left: Occupation number spectrum in $d=2$ spatial dimensions at different times. The double-log scale exhibits the bimodal power-law, with $n(k)\sim k^{-3.5}$ at low momenta. 
Center: Modulus of the field, $\phi(\mathbf{x})$ showing worm-like regions (dark grey) of near zero field. 
Right: Charge distribution $ j_0 (\mathbf{x}) $ of the field,  showing homogeneous regions of opposite charge.}
\label{fig:charge}
\end{figure}
%

We focus on a scenario of parametric resonance in a globally $O(2)$ or, equivalently, $U(1)$ symmetric relativistic scalar field theory (mass parameter $m=0$) in $d=2, 3$ dimensions~\shortcite{Gasenzer:2011by}. 
Shortly after the resonant excitations have set in, a  spatial separation of charges occurs as shown in \Fig{charge} (right panel). 
Both, charge and anti-charge overdensities become uniformly distributed within slowly varying regions which are separated by sharp boundary walls of grossly invariant thickness. 
These walls have a character similar to topological defects and appear for  initial conditions corresponding to the parametric reheating scenario. 
One observes that the presence of charge domains coincides with the visibility of  non-thermal stationary scaling solutions, see \Fig{charge} (left panel), as discussed before within the context of ultracold atomic gases. 
In this way, a link is established between wave turbulence phenomena as discussed, e.g., by \shortciteNP{Micha:2002ey,Micha:2004bv,Berges:2008wm,Berges:2008sr,Scheppach:2009wu,Berges:2010ez,Carrington:2010sz}, and long-lived quasi-topological structures in the inflaton field. 
Strong non-thermal stationary scaling solutions have also been observed for the case of $O(4)$- and $O(10)$-symmetric scalar fields in $d=3, 4$ dimensions~\shortcite{Berges:2008wm,Berges:2010ez}.
The latter work raises the interesting question after the corresponding spatial configuration at the fixed point.

\section{Outlook}

In these notes we have pointed out the possibility of a universal duality between decaying topological defects and a non-perturbative inverse wave-turbulent cascade. 
This cascade requires the generation of (quasi-)topological configurations far from thermal equilibrium and their slow decay, going together with an increase of coherence and defect separation. 
Under these conditions, we expect power-law scaling in a regime between the scales $1/\xi$, $\xi$ being the microscopic extent of the defect core, and $1/l_{\mathrm{COH}}$, where $l_\mathrm{COH}$ is the coherence length approximated by the mean distance between defects. 
In this setting, an inverse particle cascade is generated by defect dilution, and the associated power-laws can be found from the scaling properties of the respective single defect. 
We have shown this mechanism to exist in soliton- and vortex dominated single-component Bose gases~\shortcite{Nowak:2010tm,Nowak:2011sk,Schmidt:2012kw,Nowak:2012gd}, decaying domain walls and vortices in two-component Bose gases~\shortcite{Karl2012a}, as well as transient charge domains in complex relativistic scalar theory~\shortcite{Gasenzer:2011by}. 
It is emphasised that the stability of these defects does not need to be topological, as the examples of solitons and charge domains indicate. 
Cf., e.g., \shortciteN{Lee1992a} for a review of non-topological solitons.
A list of expected scaling properties is given in Table~\ref{Table}.

A variety of (quasi-)topological excitations are known to exist in superconductors, magnets, and cosmic fields~\shortcite{Lee1992a,Thouless1998a,Pismen1999a,Nelson2002a}. 
Specific examples are monopoles in gauge fields~(\shortciteANP{Rajantie2002a}, \citeyearNP{Rajantie2002a}; \citeyearNP{Rajantie2012a}) and exotic magnets~\shortcite{Castelnovo2008a}, as well as skyrmions in Bose-Einstein condensates~\shortcite{Ruostekoski2001a,Kasamatsu2005a} and liquid crystals~\shortcite{Dierking2003a}.
Coherent polariton ensembles represent a promising new route to study the dynamics of defects and solitary waves \shortcite{Amo2011a}. 

\begin{table}
\center
\renewcommand{\arraystretch}{1.3}
{\small
\begin{tabular}{|c|c|c|c|c|}
\hline
 (Quasi-) topological defect & Field & $d$ & Momentum scaling  \\
\hline
\hline
 Soliton/Domain & $\phi \in \mathbb{C},S_z \in \mathbb{R} $ & 1 & $\langle|\phi(k)|^2 \rangle \,,\, \langle|S_z(k)|^2\rangle \sim k^{-2}$ \\
\hline 
 Soliton line/Domain & $\phi \in \mathbb{C},S_z \in \mathbb{R} $ & 2 & $\langle|\phi(k)|^2  \rangle \,,\, \langle|S_z(k)|^2\rangle  \sim k^{-3}$ \\
\hline 
 Soliton surface/Domain & $\phi \in \mathbb{C},S_z \in \mathbb{R}$ & 3 & $\langle|\phi(k)|^2 \rangle \,,\, \langle|S_z(k)|^2\rangle \sim k^{-4}$ \\
\hline 
 Vortex &$\phi \in \mathbb{C} $ & 2 & $\langle |\phi(k)|^2 \rangle \sim k^{-4}$ \\
\hline
 Vortex line &$\phi \in \mathbb{C}$  & 3 & $\langle|\phi(k)|^2 \rangle \sim k^{-5}$ \\
\hline  
 Skyrmion & $\mathbf{S}\in  \mathbb{R}^3$ & 2 & $\langle|S_{x(y)}(k)|^2  \rangle \sim k^{-2}$ \\
\hline 
 Skyrmion line & $ \mathbf{S}\in  \mathbb{R}^3$ & 3 & $\langle|S_{x(y)}(k)|^2  \rangle \sim k^{-3}$ \\
\hline 
 Monopole & $ \mathbf{E}\in  \mathbb{R}^2$ & 2 & $\langle|E(k)|^2  \rangle  \sim k^{-2}$ \\
\hline 
 Monopole & $  \mathbf{B}\in  \mathbb{R}^3$ & 3 & $\langle|B(k)|^2 \rangle  \sim k^{-2}$ \\
\hline 
\end{tabular}
}
\vspace*{0.5cm}
\caption{Strong wave turbulence scaling of momentum spectra of different distributions as expected at a non-thermal fixed point, for different types of defects (first column) in various systems (second column) in $d$ dimensions.
Concerning the definitions of the defects we refer to \protect\shortcite{Rajaraman1982a,Kasamatsu2005a}.
The `Skrymion' denotes the respective defect arising in the non-linear sigma model (Anderson-Toulouse vortex).}
\label{Table}
\end{table}

The study of multi-component fields is certainly among the most interesting new directions of research in this context. 
We have taken a first step in this direction by investigating the two-component Bose gas~\shortcite{Karl2012a}. 
The possibility of different non-thermal fixed points depending on inter- and intraspecies couplings opens a perspective on new types of experiments far from equilibrium~\shortcite{Kasamatsu2006a,Nicklas2011a}. 
Relating (quasi-)topological field configurations known from equilibrium spinor Bose gases~\shortcite{Ueda2012a} to transient scaling phenomena observed in correlation functions provides a great challenge for experiments and theory. 
Experimental studies of ultracold spin-1 and spin-2 Bose gases, including the detection of spin domains, are far developed~\shortcite{Miesner1998a,Schmaljohann2004a,Chang2004a,Higbie2005a,Sadler2006a,Guzman2011a}.
Multi-component fields are important far beyond ultracold atomic physics. 
For example, multi-component inflatons and their associated topological defects are discussed in early-universe expansion scenarios~\shortcite{Rajantie2003a,Berges:2004yj,Allahverdi:2010xz}. 
Exciting recent developments in the field of heavy-ion collisions, where the non-equilibrium regime of a quark-gluon plasma can be investigated by multi-component gauge field simulations~\shortcite{Arnold:2005ef,Berges:2008mr,Carrington:2010sz,Fukushima:2011nq,Fukushima:2011ca,Berges:2012us,Berges:2012ev} may exhibit a strong relevance of the type of interrelations discussed above. 
Defect-induced non-thermal fixed points in this system are presently being explored. 
The rapid expansion of the quark-gluon plasma adds a completely new aspect to the dynamical description. 
Similar processes can be studied by releasing ultracold gases from their trapping potential, specifically in view of expanding turbulent clouds~\shortcite{Henn2009a,seman2011a,Caracanhas2012a,Weckesser2012a}. 

Ultimately, non-thermal fixed points have to be included into a global picture of non-equilibrium dynamics of interacting many-body systems~\shortcite{Polkovnikov2011a,Gasenzer2009a}. 
The concept behind them points out a way towards universal phenomena far away from equilibrium, having many aspects in common with universality and critical phenomena in thermal equilibrium. 
To understand their relations to non-thermal equilibrium states~\shortcite{Rigol2007a,Eckstein2008a,Kollar2008a,Rigol2009a,Kronenwett:2010ic}, generalised Gibbs states~\shortcite{Rigol2007a,Eckstein2008a,Kollar2008a}, or prethermalised states~\shortcite{Berges:2004ce,Bonini1999a,Aarts2000a,Barnett2011a,Kitagawa2011a,Gring2011a,Kollar2011a,Werner2012a,Tsuji2012a} are essential steps towards a unifying framework of complex dynamical many-body systems.
To set up this framework, the development and extension of renormalisation-group techniques for far-from-equilibrium dynamics seems in order, and promising progress has been seen in the recent past (see Refs.~given at the end of \Sect{IRScalingAsSWT}).
\subsection*{Acknowledgements}
{
The authors thank B.~Anderson, V.~Bagnato, J.~Berges, N.~G.~Berloff, E.~Bodenschatz, R.~B\"ucker, L.~Carr, M.~J.~Davis, S.~Diehl, B.~Eckardt, G.~Falkovich, S.-C.~Gou, H.~Horner, R.~Kerr, G.~Krstulovic, S.~Mathey, I.~Mazets, L.~McLerran, A.~Millis, M.~K.~Oberthaler, J.~M. Pawlowski, N.~Philipp, M.~G.~Schmidt, J.~Schmiedmayer, B.~Shivamoggi, B.~Svistunov, M.~Tsubota, and P.~Weckesser for useful discussions. 
T.~G. and D.~S. thank J. Berges and C. Scheppach for collaboration on related work. 
They acknowledge support by the Deutsche Forschungsgemeinschaft (GA 677/7,8), by the University of Heidelberg (FRONTIER, Excellence Initiative, Center for Quantum Dynamics), by the Helmholtz Association (HA216/EMMI), by the University of Leipzig (Grawp-Cluster), by BMBF and MWFK Baden-W\"urttemberg (bwGRiD cluster).
The authors thank KITP, Santa Barbara, for its hospitality.
This research was supported in part by the National Science Foundation under Grant No.~PHY05-51164.
T.~G.~thanks l'Ecole de Physique des Houches for its hospitality.
}


\end{document}